\documentstyle[11pt]{article}

\newcommand{\Integer}{{\bf Z}}
\newcommand{\Complex}{{\bf C}}
\newcommand{\One}{{\bf 1}}

\newcommand{\intl}{\int\limits}
\newcommand{\suml}{\sum\limits}

\newcommand{\be}{\begin{equation}}
\newcommand{\ee}{\end{equation}}
\newcommand{\ba}{\begin{eqnarray}}
\newcommand{\ea}{\end{eqnarray}}
\newcommand{\pa}[1]{\left({#1}\right)}
\newcommand{\br}[1]{\left[{#1}\right]}

\title
         {\Large\bf Rigged Hilbert Space Approach \\
to Spectral Analysis
         of the  Frobenius--Perron Operator \\ for the
         Tent--map}

\author
         {L. A. Dmitrieva, D. D. Guschin, Yu. A. Kuperin\\
         \it Laboratory for Complex Systems,
         St Petersburg State University,\\
         \it 198504 St.Petersburg, Russia\\
             e-mail: kuperin@jk1454.spb.edu}

\date{}

\begin{document}

\maketitle

{\small
   On  the  basis of an unified  theoretical formulation  of
resonances  and  resonance  states  in  the  rigged  Hilbert
spaces  the  spectral  analysis  of  the   Frobenius--Perron
operators corresponding to the exactly solvable chaotic  map has
been developed. Tent map as the simplest representative of exactly
solvable chaos have been studied in details in frames  of the
developed approach.   An  extension   of  the
Fro\-be\-nius--Perron operator resolvent to a suitable  rigged
Hilbert space has been constructed in particular and the
properties of the generalized  spectral decomposition  have  been
studied. Resonances  and  resonance projections for this map have
been calculated explicitly. }

\vspace{1cm}

        \section {Introduction}

   The statistical description of the dynamical systems in
terms of ensembles have been introduced by Gibbs and Einstein. It
allowed to calculate "suitably" the averages and, what is more
important, the notion of the ensembles proved necessary for the
description of the thermodynamical equilibrium. In general, one
can understand the thermodynamical properties systems in terms of
ensembles but not in the terms of the
trajectories~\cite{8},~\cite{27}. So, the distribution of the
probabilities becomes the main value. Then the question appears: to
which functional space the probability should belong. E.~g. for
the Bernoulli shift one can construct the spectral representation
in the Hilbert space $ L_2 (0,1) $ \cite{9}. But the spectrum of
the Bernoulli shift which is isomorphic to the unilateral
shift~\cite{6} is not linked in $ L_2 (0,1) $ with
Lyapunov time and, so, does not allow to describe the approach to
equilibrium. If we examine the Frobenius--Perron operator
\cite{7},~\cite{15} for the Bernoulli shift not in $ L_2 (0,1) $
but in the rigged Hilbert space
        $ \Phi \subset L_2 \subset \Phi^\times $
one should change the notion of the spectral
representation~\cite{10},~\cite{14}. The structure of the rigged
Hilbert space
        $ \Phi \subset {\cal H} \subset \Phi^\times $
allows to include into the spectral decomposition the strong
irregular functions which do not belong to $ L_2 (0,1) $.
In the papers on the spectral analysis of the Frobenius--Perron
operator in the rigged spaces~\cite{1},~\cite{4},~\cite{21} it
was noted that the "non-Hilbert" spectrum is linked directly
with Lyapunov time and so characterizes the temporary horizon
of the chaotic systems.

        It should be noted that the trajectories for the
Bernoulli shift can be represented by the
delta-function $ \delta (x - x_n) $ where $ x_n = S^n x_0 $
and, formally, the Frobenius--Perron operator for the
Bernoulli shift can act on the delta-function too.
Indeed, since the action to the densities is given by
the formula~\cite{6}
        \be \label {B1}
         (U \rho) (x) = \frac12 \br{ \rho \pa{ \frac{x}2 }
        + \rho \pa{ 1 + \frac{x}{2} } }
        \ee
so
        $ U: \delta (x-x_n) \rightarrow \delta(x - 2x_n) $
at
        $ x_n \le 1/2 $
and
        $ U: \delta (x-x_n) \rightarrow \delta(x+1-2x_n) $
at
        $ 1/2 \le x \le 1 \, .$
But the spectral representation in the rigged Hilbert space
        $ \Phi \subset {\cal H} \subset \Phi^\times $
        \be \label {B2}
         U = \sum_n \frac12 | B_n \rangle \langle
         \tilde B_n |
        \ee
is applicable only to smooth densities $ \rho \in \Phi $. It is
because the left eigenvector $\langle \tilde B_n |$ is some
functional of delta-like type. So, the spect\-ral
re\-pre\-sen\-ta\-tion for $ U $ in
        $ \Phi \subset {\cal H} \subset \Phi^\times $
is correct only for the smooth set of the trajectories but not for
the individual trajectory separately. It is the fundamental result
which states that for the chaotic systems the description on the
language of the distributions can not be reduced to the
description on the terms of the trajectories. It is the principal
difference of the approach based on the spectral representation in
rigged Hilbert space from the theory of the Gibbs--Einstein.

   In general, one should use the
different approaches for constructing the spectral representation
for one or another class of the operators . So, e.~g. for the
self-adjoint, unitary or normal operators the classical spectral
theorem works good. For the operators more "complicated" than the
above--mentioned something more suitable should be developed. For
the arbitrary compact operators, in particular, the spectral
theorem can not give the exhaustive description~\cite{11}. The
typical contr-example is the operator
        $ (T\rho)(x) = \intl_0^x \rho (x') \, dx' $
in $ L_2 (0,1). $ The operator $ T $ has no
eigenvalues and its spectrum is localized at zero: $ \sigma (T) = \{ 0 \} $.
The reconstruction of the spectral theory of such type of operators
needs  approaches different from
the spectral theorem~\cite{29}.

   In present paper we use the approach of the rigged Hilbert
spaces~\cite{13},~\cite{14} and construct the generalized
spectral decomposition of the Frobenius--Perron operator for the
tent--map based
on the extensions of its resolvent. This approach has been
proposed in~\cite{1} and applied, in particular, for
constructing of the spectral analysis of the Frobenius--Perron
operator for the Renyi map. In~\cite{1},~\cite{4} it has been
noted that the generalized spectrum for this map consists of
the points
        $ z_n = \beta^{-n}, \, n \ge 1, \, \beta \ge 2 $
and the left and right eigenfunctions
$ \langle \tilde \psi_n | $ and  $ | \psi_n \rangle  $
belong to  $ \Phi^\times $ and  $ \Phi $ respectively. So, the
different modes of $ \rho (x) $ approach to the equilibrium
distribution $ \rho_* (x) $
with the characteristic Lyapunov times
        $ \gamma_n = n {\rm ln} \beta $ .
This fact follows directly from the spectral decomposition
for the corresponding
Frobenius--Perron operator~\cite{1},~\cite{4}:
        \be \label {B3}
         \pa {U^t_R \rho}(x) = \sum_{n \ge 0}
         e^{- \gamma_n t} | \psi_n \rangle
         \langle \tilde \psi_n | \rho \rangle \, ,
         \quad t = 1, \, 2, \, 3\, \ldots \quad .
        \ee

   As it mentioned above we
study the dynamic of the chaotic tent map
        \be \label {B4}
         S: \,\, x \to
         \left \{
         \begin{array}{ll}
        2x,   & 0\le x \leq 1/2            \\
        2-2x, & 1/2 \le x \le 1
       \end{array}
         \right.
        \ee
on the base of the spectral representation of its
Frobenius--Perron operator $ U_T $ in the suitable rigged
space
        $ \Phi \subset L_2 \subset \Phi^\times $.
So, instead of studying individual trajectories,
generating by the iterations of the map $ S $:
        $ x_n = S^n x_0 \, , \quad x_0 \in [0,1] $
we use an approach characteristic for the
nonequilibrium statistical mechanics [12], i.e. we observe the
evolution in the time of the densities
        $ \rho (x) \in \Phi .$
The evolution of the latter is given by
the operator
        $ U^t_T \, , \quad t = 1, \, 2, \, 3\, \ldots \, , $
i.e.  $ \rho (x,t) = U^t_T \rho_0 (x) $,
where $ \rho_0 (x) $ is an initial density of the distribution.
In our approach the rate of the decay of the
initial density is determined by the non--Hilbert spectrum
        $ \sigma (U^t_T) $
of the Frobenius--Perron operator. In frames of the
approach developed in~\cite{24} the rates of the
decay correspond to the poles of the Fourier
transformation of the correlation function and so, are
interpreted as the resonances of the dynamical system. We
note that for some dynamical systems these poles has been
studied in~\cite{30},~\cite{31},~\cite{32} on the basis of the
theory of the periodical orbits and dynamical
$ \xi $-function~\cite{22},~\cite{23}. So, the
generalized spectrum in our approach coincides with what is
known as Ruelle--Pollicott resonances.

   The detailed operator construction, using in this work,
is described briefly in following section. Here we note only that
our approach is similar to the methods of~\cite{4},~\cite{1},
\cite{16},~\cite{25} and is different essentially from the
technique used e.g. in~\cite{22}.

        \section {Spectral decomposition in rigged
                  Hilbert \- spaces}

   In  this section we introduce necessary notions
and describe briefly the approach based on the
extensions of the Frobenius--Perron operator for
one-dimensional maps.

   Let
        $ S: [0,1] \to [0,1] $
be one-dimensional map,  $ x \in [0,1] $. Denote as $ S^{-1}
([0,x])  =  \{ y: \,\, S(y) \in [0,x] \} $  and let $ \rho (x) $
be a density on $ [0,1] $ . We suppose that $\rho (x)$ is so, that
all integrals below exist. Then the Frobenius--Perron operator $U$
associated with the map $S$ is given by the
formula~\cite{7},~\cite{15}
        \be \label {c1}
         \int_0^x U \rho (x') \, dx' =
         \int_{S^{-1} ([0,x]) } \rho (x') \, dx'.
        \ee
The Koopman operator $ V $ for the map $ S $ can be defined
as~\cite{7},~\cite{15}
        \be \label  {c2}
         ( V \rho ) (x) = \rho (Sx) .
        \ee
We shall start with the densities $\rho$ such that
        \be \label {c3}
         \rho \in {\cal H} = L^2(0,1) .
        \end{equation}
Then, denoting inner product in $ {\cal H} $ as
        $<\cdot \;,\; \cdot>$
using~(\ref{c1}),~(\ref{c2}) one can show that for any
functions $f$ and $g$ from $ {\cal H} $,
        \be \label {c4}
         \langle Uf , g \rangle = \langle f , Vg \rangle .
        \ee
So, the Koopman operator $V$ is adjoin to Frobenius--Perron
operator $U$, i.~e.
        \be \label {c5}
         V = U^* .
        \ee
Equations~(\ref{c4}) and~(\ref{c5}) give a possibility to
calculate $U$ if we know $V$ and vice verse. For the tent--map one
can obtain
        \be \label{c6}
         S^{-1} ([0, x]) = \br{ 0, \frac{x}2 } \cup
         \br{ 1 - \frac{x}2, 1 }
        \ee
and
        \be \label{c7}
         \int_0^x U_T \rho (x') \, dx' = \int_0^{ x/2 }
         \rho (x') \, dx'+ \int_{ 1-x/2 }^1 \rho (x') \, dx'.
        \ee
It means that $U_T$ reads as
        \be \label {c8}
         (U_T \rho) (x) = \frac12
           \br{ \rho \pa{ \frac{x}2 }
         + \rho \pa{ 1 - \frac{x}2 }}.
        \ee
On the other hand, according to~(\ref{c4}),~(\ref{B4}) one have
        \begin{eqnarray*}
         \langle V_T f, g \rangle \!\!\!\! &=& \!\!\!\!
         \int_0^1 f (Sx) \overline{ g(x) } \, dx =
         %\\  &=&
         \int_0^{1/2} f (2x) \overline{ g (x) } \, dx +
         \int_{1/2}^1 f ( 2(1-x) ) \overline{ g (x)} \, dx =\\
         &=& \!\!\!\! \frac12 \int_0^1 f (x) \overline{ g \pa{\frac{x}{2}}} \, dx
         + \frac12 \int_0^{1/2} f (x)
         \overline{ g \pa{1-\frac{x}{2}}} \, dx =
         \\ &=& \!\!\!\!
         \int_0^1 f (x) \frac12
         \br{ \overline{ g \pa{\frac{x}{2} }} +
         \overline { g \pa {1 - \frac{x}{2} } } } \, dx =
         \langle f, V_T^* \rangle = \langle f, U_T g \rangle \, ,
        \end{eqnarray*}
that leads again to the equation~(\ref{c8}).

   In terms of Frobenius--Perron operator it is possible
to determine the stationary (equilibrium) density $ \rho_* (x) $
and corresponding invariant measure $ \mu_* (x) $.
Namely $ \rho_* (x) $  is determined by
the equation
        \be \label {c9}
         U \rho_* = \rho_* ,
        \ee
and $ \mu_* (x) $ is given by the formula
        $$
         \mu_* (x) = \int_0^x \rho (x') \, dx' .
        $$
Solving the functional equation for the
equilibrium density $ \rho_* $
        \be \label {c10}
         \rho_* (x) = \frac12
         \br{ \rho_* \pa{ \frac{x}2 }
         + \rho_* \pa{ 1 - \frac{x}2 } }
        \ee
one can find that
        \be \label {c11}
         \rho_* (x) = 1
        \ee
and hence
        \be \label {c12}
         \mu_*(x) = x .
        \ee
So the invariant measure of the tent--map is the Lebegues measure.

   We shall study the operators  $ U_T $  and  $ V_T $  in the
rigged  Hilbert  spaces and  in  this connection,  we  shall
describe  the scheme of  the spectral analysis in these
spaces~\cite{1} below.

   The space  $ \Phi $ is a test space for
the operator $ U_T $ if:

   1) $ \Phi $ is a locally convex topological vector space
with  topology  stronger  than the  Hilbert space topology
${\cal H}$;

   2) $ \Phi $  is  continuously  and  densely  embedded in
$ {\cal H} $;

   3) $ \Phi $ is invariant to the adjoin operator $ U^*_T $:
i.~e. to $V_T$: $ U^*_T~\Phi~\subset~\Phi $.

   The dual space $\Phi^\times$ to $ \Phi $ we shall consider
as the linear continuous functionals on $\Phi$. It is clear, that
the topology in $\Phi^\times$ is weaker that in  $ \cal H $. The
space $\Phi \subset {\cal H} \subset \Phi^ {\times}$ we shall call
rigged Hilbert space or Gelfand triplet. The coupling in
    $\Phi \subset {\cal H} \subset \Phi^{\times}$
we shall denote as $<\cdot\,| \,\cdot>$.

   The operator $\tilde U_T$ is called an extension of the
operator $U_T$ to the dual space
$\Phi^{\times}$ if $\tilde U_T$ acts on linear functionals
$\langle f | \in \Phi^{\times}$ as:
        \be \label {c13}
         \langle \tilde U_T f | \varphi \rangle =
         \langle f | U^*_T \varphi \rangle
        \ee
for all test functions $ \varphi \in \Phi $.

   We define the extended resolvent $\tilde R_U(z)$ to a
rigged Hilbert space as follows~\cite{1}: we shall call
the operator--valued function $\tilde R_U(z)$ an extended resolvent
of the operator $U$ in a suitable rigged Hilbert space
        $\Phi \subset {\cal H} \subset \Phi^{\times}$
if:

   1) $\tilde R_U(z)$ satisfies the resolvent identity in
the weak sense:
      \be \label {c14}
       \langle \varphi | \tilde R_U(z)(U_T-z) | \psi \rangle
       = \langle \varphi | \psi \rangle ,
         \quad \forall \varphi, \psi \in \Phi \, ,
      \ee
where z is not a singular point of the $\tilde R_U(z)$;

   2) $\tilde R_U(z)$ satisfies a completeness condition in
a weak sense, i.~e.
        \be \label {c15}
         - \frac1{2\pi i} \oint_{\Gamma} \langle \varphi |
         \tilde R_U(z) | \psi \rangle dz\, =
         \langle \varphi | \psi \rangle ,
         \quad \forall \varphi, \psi \in \Phi \, ,
        \ee
where the contour $\Gamma$ encircles all the singularities of the
integrand in~(\ref {c15}) in the positive direction.

Then,
from~(\ref{c14}) the dependence between the extended
operator $\tilde U_T$ and the extended resolvent  $\tilde
R_U(z)$, is, as usual,
      \be \label {c16}
       \tilde R_U(z)=\pa {\tilde U_T - z {\bf I}}^{-1} .
      \ee
So the extended operator $\tilde U_T$ corresponds to the extended
resolvent $\tilde R_U(z)$.

   We define the Gelfand spectrum
$\hat{\sigma}(\tilde U_T)$ of the operator $\tilde U_T$ as a set
of singularities of the extended resolvent $\tilde R_U(z)$ on the
complex plane $z\in\Complex$. We shall not give here the detailed
classification of the spectral components of $\hat{\sigma}(\tilde
U_T)$ and restrict ourselves by the case when all singularities of
the extended resolvent $\tilde R_U(z)$ are the poles:
$\hat{\sigma}(\tilde U_T)=\{z_\nu\}$. The residues in the poles
$\{z_\nu \}$ of $\tilde R_U(z)$ will be referred as the
generalized spectral projections of the operator $\tilde U_T$:
        \be \label {c17}
         \Pi_{\nu}={\rm res} \tilde R_U(z)|_{z=z_{\nu}} =
         - \frac1{2\pi i} \oint_{C_{\nu}} \tilde R_U(\zeta)
         d\zeta \,.
        \ee
For the case of simple poles $\Pi_{\nu}$ is one-dimensional
projection:
        \be \label {c18}
         \Pi_{\nu} = | \psi_{\nu} \rangle \langle \tilde
         \psi_{\nu} | \, ,
        \ee
where $|\psi_{\nu} \rangle$ and $\langle \tilde \psi_{\nu}
|$ are the generalized right and left eigenvectors of the
corresponding extended operator $\tilde U_T$:
        \ba \label {c19}
         \tilde U_T |\psi_{\nu} \rangle = z_{\nu} |\psi_{\nu}
         \rangle \,, \nonumber \\
         \langle \tilde \psi_{\nu}| \tilde U_T = z_{\nu}
         \langle \tilde \psi_{\nu}| \,\,.
        \ea
Then, from the decomposition~(\ref{c15}) it follows that the
generalized spectral decomposition for $ \tilde U_T $ looks like
        \begin{equation} \label{c20}
         \tilde U_T = \suml_{\nu}\,z_{\nu}\,\Pi_{\nu} .
        \end{equation}
The latter is considered as decomposition of density
         $ \rho \in \Phi $ by the (biorthogonal) system of
eigenfunctions
        $ \{ | \psi_\nu \rangle\,,\,\langle \tilde \psi_\nu |\} $:
        \be \label {c21}
         \tilde U_T \rho = \suml_{\nu}\,z_{\nu}
         | \psi_\nu \rangle \langle \tilde \psi_\nu | \rho \rangle \, .
        \ee

   The described construction can be used also for the extension
of the Koopman operator $\tilde V_T$~\cite{1},~\cite{16}.

   Summarizing we can formulate the following algorithm for
the construc\-tion of the spectral analysis of the Frobenius--Perron
operator in the triplet
        $ \Phi \subset {\cal H} \subset \Phi^\times $:

     1. Fixe the test space $\Phi$ in $\cal H$ and study the
restriction $R_U (z) |_\Phi$ of the resolvent of the operator $U_T$.

     2. Make the extension of the resolvent $R_U (z) |_\Phi$
to $ \tilde R_U (z) $ and find all singularities $ \tilde R_U (z)
$ in the complex plane $z \in \Complex$.

     3. Construct the decomposition of the~$ \tilde R_U (z) $
in weak sense, i.~e. verify the conditions of the
completeness~(\ref{c15}).

     4. If all singularities of $ \tilde R_U (z) $ are the
poles,  then calculate the generalized spectral projections~$
\Pi_\nu$.

     5. Study the convergence of the generalized spectral
de\-com\-posi\-tion~(\ref{c20}) in the sense of~(\ref{c21}).

     In the following sections this algorithm shall be realized
for the Frobenius--Perron operator corresponding to the tent map.

\section  {The resolvent of Frobenius--Perron operator in the
           Fourier representation}

   Let  $ {\cal A}[0,1] $ be the algebra of analytic
functions on the interval $ [0,1] $.  It is known that
$ {\cal A} [0,1] $ is continuously and densely
embedded into the space $ L_2[0,1) $ and so one can consider
the restriction of the Frobenius--Perron operator $ U_T $ on
$ {\cal A}[0,1] $ choosing the latter as the space $ \Phi $ of
the  test  functions in the Gelfand triplet  $ \Phi  \subset
L_2[0,1] \subset \Phi^\times $.

   Let $F$ be the Fourier transform on ${\cal A}[0,1]$.
Then,
      \be \label {1}
       F: {\cal A}[0,1] \longrightarrow
         \widehat{{\cal A}[0,1]}
       \subset l_2 \, ,
      \ee
      \be \label {2}
       F\rho(x) = \widehat \rho =
       \{\rho_n\}_{n=-\infty}^{n=+\infty} \in l_2 \, ,
      \ee
      \be \label {3}
       \rho(x) = \sum_{n=-\infty}^{\infty} \rho_n
       e^{2i\pi nx}, \qquad
       \rho_n = \int_0^1 \rho(x) e^{-2i\pi nx} dx \, ,
      \ee
and the series in~(\ref{3}) converges in $L_2$-norm.

   Let us calculate the action of the operator
$\hat U_T = FU_TF^{-1}$:
        \begin{eqnarray*}
         (U_{T}\rho)(x) &=& \sum_{n=-\infty}^\infty
         \rho_{n}\,U_{T}\, e^{2i \pi nx} =
         \frac12 \sum_{n=-\infty}^\infty
         \rho_{n}\, \left( e^{i \pi nx}+e^{-i \pi nx}
         \right) =\\
         &=& \frac12 \sum_{n=-\infty}^\infty
         \left( \rho_{n}+\rho_{-n}\,\right) \,e^{i \pi nx}.
        \end{eqnarray*}
Since
        $$
         (U_T\rho)(x)=\sum_{m={-\infty}}^\infty
         (\hat U \rho)_m e^{2i \pi mx} ,
        $$
the coefficients of these series are given by the expression
        \begin{eqnarray*}
         (\hat U_{T}\rho)_{m} \!\!\!\! &=& \!\!\!\!
         \int_{0}^{1} ( \hat U_{T}\rho )(x)\,
         e^{-2i \pi mx} dx %\\
         = \frac12
           \sum_{n=-\infty}^\infty
         \left(\rho_{n}+\rho_{-n}\,\right)
         \int_{0}^{1}\,e^{i \pi (n-2m)x} dx.
        \end{eqnarray*}
Dividing the latter series into two series with summation by even
and odd indices we have
        \begin{eqnarray*}
         (\hat U_{T}\,\rho)_{m}\!\!\!\! &=& \!\!\!\! \frac{1}{2} \biggl(
         \sum_{k}\left(
         \rho_{2k} + \rho_{-2k}\right)
         \int_{0}^{1}\,e^{i \pi(2k-2m)x} dx\\ &&\!\!\!\!+
         \sum_{k}\left(\rho_{2k+1} + \rho_{-(2k+1)}\right)
         \int_{0}^{1}\,e^{i \pi (2k+1-2m)x} dx \, \biggr) \, .
        \end{eqnarray*}
The first integral in the right hand side of this expression
equals to $\delta_{km}$. So we have
        \be \label {4}
         \pa{\hat U_T \rho}_m=\frac12 \pa{\rho_{2m} +
         \rho_{-2m}} + \frac{i}\pi \sum_k
         \frac{\rho_{2k+1} + \rho_{-(2k+1)}}{2k-2m+1} \,
        \ee
where we used the relation
        $$
         \int_0^1 e^{i \pi (2k+1-2m)x}dx =\frac{2i}\pi \cdot
         \frac1{2k+1-2m} \, .
        $$

    The representation(\ref{4}) can be used
for finding the so\-lu\-tion of the equation
      \be \label {5}
         \pa{ (\hat U_T - z) \rho}_m = f_m \, , \quad
         \{f_m\}_{m = - \infty}^{m=+\infty} \in l_2 \, ,
        \ee
which is the base for obtaining the resolvent
        $ \hat R_T (z) = \pa{\hat U_T - z}^{-1} $
of the operator $\hat U_T $. Indeed,  in an explicit
form~(\ref{5}) reads for $m \ne 0$ as
      \be \label {6}
         \rho_m = \frac1{z}
         \br{ \frac12 (\rho_{2m} + \rho_{-2m}) +
         \frac{i}{\pi}
         \sum_k \frac {\rho_{2k+1}+\rho_{-(2k+1)}}{2k-2m+1}}
         -\frac1{z} f_m \, ,
      \ee
and for $ m = 0 $ as
        \be \label {7}
       \rho_0 = \frac1{1-z} f_0 \, .
        \ee
It is possible to solve~(\ref{6}) by iteration, as it was done
in~\cite{1},~\cite{4} for the Renyi map. But the iteration
procedure for the equation~(\ref{6}) turns out to be very
cumbersome. Therefore we shall use an additional trick based on
the Feshbach projection technique~\cite{2} and the properties of
symmetry~\cite{3} of the map with respect to the point $x = 1/2$.

\section {The solution of the equation for the resolvent by
          the Feshbach projection method}

   Let us define for all functions $ f $,
        $ f \in {\cal A}[0,1] $
the operator $R$ of the reflection with respect to the point
$ x = 1/2 $ by the formula
        \be
         Rf(x) = f(1 - x) \, ,
        \ee
and in terms of $R$ on $ {\cal A}[0,1] $ we define the pair of
the operators
        $$
         P^{\pm} = \frac12 ({\rm I} \pm R) \, .
        $$
Obviously, that the operators $ P^\pm $ satisfy the following
equations:
        \ba \label {9}
         && P^+ + P^- = {\rm I}   \, ,   \nonumber   \\
         && (P^\pm)^2 = P^\pm     \, ,   \nonumber   \\
         && (P^+)^* = P^-         \, ,   \nonumber   \\
         && P^+ P^- = P^- P^+ = 0 \, .
        \ea
which mean that $P^+ , \,\, P^-$ is the pair of the
orthoprojections in
        $ {\cal A}[0, 1] \subset L_2 (0, 1) \, . $
Hence, using $P^+ , \, P^-$ one can split $ L_2 (0, 1) $
into the orthogonal sum $ L_2^+ \oplus L_2^- $
of the even and odd functions with respect to the point
        $ x = 1/2 $
respectively.

   Following~\cite{1} we rewrite the equation for the resolvent
        $$
         (U_T-z)\rho(x)=f(x)
        $$
in the form:
        \be \label {10}
         (U_T - z)(P^+ + P^-) \rho(x) = f(x) \, .
        \ee

  Acting by $P^+$ from the left on~(\ref{10}) one can obtain
according to~(\ref{9})
        \be \label {11}
         P^+ U_T P^+ \rho - zP^+ \rho + P^+ U_T P^- \rho = P^+ f .
        \ee
Similarly, acting on~(\ref{10}) by the operator $ P^- $ from the
left we get
        \be \label {12}
         P^- U_T P^- \rho - zP^- \rho + P^- U_T P^+ \rho = P^- f
        \ee
Denoting
        $ P^{\pm} \rho = \rho^{\pm}, \quad P^{\pm} f  = f^{\pm} $,
we obtain from~(\ref{11}),~(\ref{12}) the set of equations
with respect to the vector $ \rho^+ \oplus \rho^- $:
        \be \label {13}
         \left( \begin{array}{cc}
                 P^+\,U_T\,P^+-z & P^+\,U_T\,P^- \\
                 P^-\,U_T\,P^+   & P^-\,U_T\,P^--z
                \end{array}
         \right)  \cdot
         \left( \begin{array}{c} \rho^+ \\ \rho^- \end{array}
         \right) =
         \left( \begin{array}{c} f^+ \\ f^- \end{array}
         \right).
        \ee
Let us note that for the functions symmetrical with respect to the
point $ x = 1/2 $
        \be \label {14}
         \rho_n = \rho_{-n}
         \qquad {\rm and} \qquad
         \rho_n = - \rho_{-n}\,,\quad n \in \Integer
        \ee
for the functions antisymmetrical with respect to the point
$ x = 1/2 $.  Indeed, for the symmetrical functions
$ \rho (x) $ :
        \begin{eqnarray*}
         \rho(1-x)\!\!\!\! &=& \!\!\!\!
         \sum_{n=-\infty}^\infty\rho_{n} e^{2i \pi n(1-x)}=
         \sum_{n=-\infty}^\infty\rho_{n} e^{2i \pi n}
         e^{-2i \pi nx} =\\
         &=& \!\!\!\! \sum_{n=-\infty}^\infty\rho_{-n} e^{2i \pi nx}=
         \sum_{n=-\infty}^\infty\rho_{n} e^{2i \pi nx} =
         \rho(x) \, .
        \end{eqnarray*}
and analogously for the antisymmetrical functions $ \rho (x) $.
Then from~(\ref{14}), it follows, that
        \be \label {14,5}
         U_T P^- \rho^- = 0 .
        \ee
It means that for the solution of the system~(\ref{13}) in the
Fourier representation only $ P^+ U_T P^+ \rho^+ $, and
 $ P^- U_T P^+ \rho^+ $ blocks
should be calculated. Namely:
        \begin{eqnarray*}
         \left( P^+ U_T P^+ \right ) \rho^+ =
       P^+ U_ T \rho^+  \!\!\!\! &=& \!\!\!\!
       P^+ \sum_{n=-\infty}^{\infty}\rho_n^+
         e^{i \pi nx} \!\!\!\! =                       % \\
       \sum_{n=-\infty}^{\infty}\rho_n^+
%&=& \sum_{n=-\infty}^{\infty}\rho_n^+
       \frac{1+R}{2} e^{i \pi nx}      =   \\
       = \sum_{n=-\infty}^{\infty}\rho_{n}^{+}\,
         \frac{e^{i \pi nx}+\!\!\!\!
         e^{i \pi n}\,e^{-i \pi nx}}{2}\,
       &=& \!\!\!\! \sum_{n=-\infty}^{\infty}\rho_{n}^{+}\,
         \frac{e^{i \pi nx}}{2}\,+\,
         \sum_{n=-\infty}^{\infty}\rho_{-n}^{+}\,
         \frac{e^{-i \pi n}\,e^{i \pi nx}}{2}=\\
      &=& \!\!\!\! \sum_{n=-\infty}^{\infty}\rho_{n}^{+}\,
         \frac{1+e^{i \pi n}}{2}\, e^{i \pi nx}.
        \end{eqnarray*}
Since
        $ 1 + e^{- i \pi n} = 0 $
for
        $ n = 2m + 1 $ ,
then
        $$
         \pa{P^+ U_T P^+} \rho^+ =
         \sum_{m = - \infty}^{\infty}
         \rho_{2m}^+ e^{2i \pi mx} ,
        $$
and we find
        \be
         \pa{P^+ \hat U_T P^+ \rho^+ }_m = \rho_{2m}^+
         \qquad m \in \Integer \, .
        \ee
In the same manner
        \begin{eqnarray*}
         \left( P^- U_T P^+ \right ) \rho^+  =
       P^- U_ T \rho^+ \!\!\!\! &=& \!\!\!\!
       P^- \!\sum_{n=-\infty}^{\infty}\rho_{n}^+
         e^{i \pi nx}
      = \sum_{n=-\infty}^{\infty}\rho_{n}^{+}
       \frac{1-R}{2} e^{i \pi nx}=              \\
      = \sum_{n=-\infty}^{\infty}\rho_{n}^{+}\,
         \frac{e^{i \pi nx} -
       e^{i \pi n}\,e^{-i \pi nx}}{2} \!\!\!\!
       &=& \!\!\!\! \sum_{n=-\infty}^{\infty}\rho_{n}^{+}\,
         \frac{e^{i \pi nx}}{2}\,-\,
       \sum_{n=-\infty}^{\infty}\rho_{-n}^{+}\,
       \frac{e^{-i \pi n}\,e^{i \pi nx}}{2}=\\
       &=& \!\!\!\! \sum_{n=-\infty}^{\infty}\rho_{n}^{+}\,
         \frac{1-e^{i \pi n}}{2}\, e^{i \pi nx}.
        \end{eqnarray*}
Since
        $ 1 - e^{- i \pi n} = 0$
for
        $ n = 2m $,
so
        $$
         \pa{P^- U_T P^+} \rho^+ =
         \sum_{m = - \infty}^\infty
         \rho_{2m + 1}^+ e^{i \pi (2m+1)x}.
        $$
Hence,
        \ba
         \left( P^-\hat U_T P^+\,\rho^+\right)_{l} =
         \,\sum_{m = -\infty}^{\infty}\rho_{2m+1}^+\,
         \intl_0^1 \,e^{i \pi (2m+1)x-2i l \pi x}\,dx=
         %\\ \nonumber
          \sum_{m = -\infty}^{\infty}\rho_{2m+1}^+\,
         \frac{i}{\pi (2m+1-2l)}.
        \ea

    So the set of equations~(\ref{13}) in the Fourier
representation gets the form ( for $ m \neq 0 $ ):
        \be \label {18}
        \left\{
         \begin{array}{rcl}
          \rho^+_{2m}-z\,\rho^+_m&=&f^+_m \\
          \suml_{l = - \infty}^\infty \rho^+_{2l+1}
          \frac{2i}{\pi(2l-2m+1)}-z\rho^-_m&=&f^-_m.
         \end{array}
        \right.
        \ee
For the case $ m = 0 $ the solution $\rho_0$ is given by the
formula~(\ref{7}).

    Let us note that in~(\ref{18}) only the first
equation for $ \rho_m^+ $ is essential because if $ \rho_m^+ $ is
founded, then $ \rho_m^- $ is reconstructed from the second
equation of the set:
        \be \label {19}
         \rho_m^- = \frac1z \suml_{l = - \infty}^\infty
         \rho^+_{2l+1}
         \frac{i}{\pi(2l-2m+1)} -\frac1z f^-_m \, .
        \ee
On the other hand, the first equation of the set~(\ref{18})
        \be \label {20}
         \rho^+_{2m}-z\,\rho^+_m=f^+_m
        \ee
coincides with the equation for the
resolvent of the Frobenius--Perron operator corresponding to
the Bernoulli shift~\cite{1} and its solution is given by the
formula (for $m \neq 0$)
        \be \label {21}
         \rho^+_m=-\frac1{z} \suml_{k=0}^\infty
         z^{-k} f^+_{2^k\,m} .
        \ee
The series in~(\ref{21}) converges absolutely if
$|z| > 1/2 $,~\cite{1} and
        $\hat{f} \in \hat{\cal A}[0,1] \subset l_2 $.
In terms of $\rho^+_m$ one can reconstruct $\rho^+_m$:
        \be \label{22}
         \rho_m^- = \frac1{z^2} \suml_{l = - \infty}^\infty
         \frac{i}{\pi(2l-2m+1)} \suml_{k = 0}^\infty
         z^{-k} f_{(2l+1)2^k}^+ - \frac1z f^-_m \, .
        \ee

   Taking into account~(\ref{14,5}) we can rewrite~(\ref{13}) as
        \begin{eqnarray*}
        \left\{
         \begin{array}{l}
          \left( P^+ U_T P^+ - z \right) \rho^+ = f^+   \\
           P^- U_T P^+ \rho^+ - z \rho^- = f^- .
         \end{array}
        \right.
        \end{eqnarray*}
Then
        \begin{eqnarray*}
        \left\{
         \begin{array}{l}
          \rho^+ = \left( P^+ U_T P^+ - z \right)^{-1} f^+ \\
          \rho^- = \frac1{z} \left( P^- U_T P^+ \rho^+ \right)
          - \frac1{z} f^- .
         \end{array}
        \right.
        \end{eqnarray*}
Hence,
        \be \label {23}
         \begin{array}{c}
         \rho^+ + \rho^- =
         P^+\rho + P^-\rho = \rho = \\
         = \br{ \pa{P^+ U_T P^+ - z}^{-1} P^+ +
         \frac1{z} P^- U_T P^+ \pa{ P^+ U_T P^+ - z }^{-1}P^+
         - \frac1{z} P^- } f \, .
        \end{array}
        \ee

   Comparing~(\ref{23}) with the presentation of the
resolvent for the ope\-ra\-tor~$ U_T $:
        $$
         \rho (x) = (U_T - z)^{-1} f \, ,
        $$
we find that the complete resolvent $(U_T - z)^{-1}$ can be
reconstructed from the partial resolvent
        $ \pa{P^+ U_T P^+ - z}^{-1} $
by the formula
                         %\ba \label {24}
                         % (U_T -z)^{-1} &=& \pa{P^+ U_T P^+ - z}^{-1} P^+
                         % \nonumber \\ &&+
                         % \frac1{z} P^- U_T P^+ \pa{ P^+ U_T P^+ - z }^{-1}P^+
                         % - \frac1{z} P^- .
                         %\ea
        \be \label {24}
         (U_T -z)^{-1} = \pa{P^+ U_T P^+ - z}^{-1} P^+
         \frac1{z} P^- U_T P^+ \pa{ P^+ U_T P^+ - z }^{-1}P^+
         - \frac1{z} P^- .
        \ee

   It is obvious that
        $
         f_n^\pm = \frac12(f_n \pm f_{-n}) \, .
        $
Then from~(\ref{21}) and~(\ref{22}) we find for
        $ n \ne 0 $
the Fourier presentation for~(\ref{24}):
        \ba \label {four}
         \rho_n = \rho^+_n + \rho^-_n =
         - \frac1{z} \suml_{k=0}^\infty
         \frac{f_{n2^k}+f_{-n2^k}}{2z^k}
         -\frac1{z} \frac{f_n -f_{-n}}{2} %\nonumber \\
         - \frac{i}{2\pi z^2}
         \suml_{l=-\infty}^\infty \suml_{k=0}^\infty
         \frac{f_{(2l+1)2^k}+f_{-(2l+1)2^k}}{z^k(2l-2n+1)}.
        \ea

   Let us show that both series in~(\ref{four}) converges
absolutely in $ z $ if $ |z| > 1/2 $. Since
        $ \rho (x) \in {\cal A}[0,1] $
then integrating~(\ref{3}) by parts we obtain:
        \ba
         &&\left. \left. \left. \rho_{m}=\intl_{0}^{1} \rho(x) e^{-2i \pi mx}dx =
         \frac{i}{2\pi m} \right[ \rho(x)\right|_{0}^{1} -
         \intl_{0}^{1} \rho'(x) e^{-2i \pi mx} dx \right] =
         \nonumber \\
        && \qquad \qquad \quad = \frac{i}{2\pi m}\left[\rho(1)-\rho(0) -
         \int\limits_{0}^{1} \rho'(x) e^{-2i \pi mx} dx \right] ,
        \ea
where $ \rho' (x)$ denotes the derivative. Denoting
        \be
         c_\rho = (2 \pi)^{-1}
         \left(
         |\rho (1) - \rho (0)| + \int_0^1 |\rho'(x)| \, dx
         \right) \, ,
        \ee
we are convinced that
         $ 0 < c_\rho < \infty $
and
         \be
         |\rho_m| \le \frac{c_\rho}m \, .
         \ee
Since
        $ \hat f \in \widehat{ {\cal A} [0,1]} $
then analogously we obtain that
        $$
         |f_m| \le \frac{c_f}m
        $$
where $c_f$ is some constant.
 It means that at $k \to \infty$ the following
estimation
        \be
         \left|\frac1{2z} f_{m2^k} \right| \le
         c_fm^{-1} \pa{2|z|}^{-k}
        \ee
is valid. Hence the first series in~(\ref{four}) converges
absolutely at
        \be
       |z| > \frac12 \, .
        \ee
Obviously, this condition is enough for the absolute
convergence of the se\-cond series in~(\ref{four}).

   Taking into account~(\ref{four}) we obtain the matrix elements
        $ \hat R_{mn}(z) $
of the resolvent of the Frobenius--Perron operator
        $ U_T $
for the tent map as (for $ n \ne 0 $):
        \ba \label {31}
         \hat R_{nm}(z) =\!\!\!\!&-& \!\!\!\!\frac12 \suml_{k=0}^\infty
         \frac1{z^{k+1}}
       (\delta_{2^kn,\,m}+\delta_{-2^kn,\,m})-
       \frac1{2z}(\delta_{n,m}+\delta_{-n,m}) \nonumber \\
         &-&\!\!\!\! \frac12 \suml_{k=0}^\infty \frac{i}{z^2 z^k \pi}
         \suml_{l=-\infty}^\infty
         \left[
         \frac{\delta_{(2l+1)2^k, m}+
         \delta_{-(2l+1)2^k, m}}{2l-2n+1}
       \right] \, ,
        \ea
and for $ n = 0 $:
        \be \label {Roo}
         \hat R_{00}(z) = \frac1{1-z}
        \ee

   From~(\ref{31}),~(\ref{Roo}) it follows that the
resolvent
        $ \hat R(z) = \{ R_{mn} \} $
is the mero\-morphic function in the domain
        $ |z| > 1/2 $
with a single simple pole at
        $ z = 1 $.
Obviously the same is valid for the resolvent
        $ R(z) = (U_T - z)^{-1} = F^{-1} \hat R(z) F $
in the space
        $ {\cal A} [0,1] $.

The residue  of the resolvent
        $ \hat R(z) $
at the point
        $ z = 1 $
is given by the integral
        \be
         {\rm res} \hat R(z)|_{z = 1} =
         - \frac1{2i \pi} \oint_\Gamma \hat R(z) \, dz ,
        \ee
where the contour $ \Gamma $ is given by
        $$
         \Gamma = \{z: z = 1 + re^{i \varphi}, \quad
         r \in (0, 1/2), \quad \varphi \in [0, 2\pi) \} \, .
        $$
In virtue of the analyticity of $ \hat R(z) $ only~(\ref{Roo})
gives the contribution in the residue. Then, denoting by
        $ \tilde \Pi_0 $ the operator with  the matrix elements
        $ \tilde \Pi_{nm}~=~\delta_{n,0} \delta_{m,0} $,
we see that
        \be
         {\rm res} \hat R(z)|_{z = 1} = \tilde \Pi_0 \, .
        \ee

Let us note, that introduced operator $ \tilde \Pi_0 $
is the projection on the vector
        $ \hat q $,
        $ q_n = \delta_{n,0} $.
In the space
        $ {\cal A}[0,1] $
the corresponding operator
        $ \Pi_0~=~F^{-1} \tilde \Pi_0 F $,
        \be
         F^{-1} \tilde \Pi_0 F = {\rm res} R(z)|_{z = 1}
        \ee
has the kernel
        \be
         \Pi_0 (x,x') =
         \sum_{n,m} \tilde \Pi_{nm} e^{2i \pi nx}e^{-2i \pi mx'}
         = \One(x) \One(x) \, ,
        \ee
where
        $ \One(x) $
is the function from
        $ {\cal A}[0,1] $
which identically equals to one on the interval
        $ [0,1] $.

    Using bra- and ket- Dirac notions, the action of the
operator
        $ \tilde \Pi_0 $
one can write formally as
        \ba \label {37}
         \Pi_0 : \,\, \rho (x) \to (\Pi_0 \rho) (x) =
         < \One(x) | \, \rho(x) > \One(x) =
         \int\limits_0^1 \rho(x) \, dx \cdot  \One(x).
        \ea

\section {The extension of the resolvent and the
          ge\-ne\-ra\-lized spectral decomposition for the
          tent map}

   According to the general approach~\cite{1} one should
extend the resolvent of the tent map outside its own domain of
analytically
        $ C_{1/2} = \{z: \,\, |z| > 1/2 \} $
i.e. into the circle
        $ C_{1/2} $.
According to~\cite{1} the extension of the
resolvent into this circle is possible in the topology
weaker than Hilbert topology.

   To build this extension, at first we calculate the kernel
        $ R(x,x',z) $
of the resolvent
        $ R(z) = (U_T - z)^{-1} = F^{-1} \hat R(z) F $ \,
in the initial space
        $ {\cal A}[0,1] .$
This kernel is given by
        \be
         R(x,x',z) = \sum_{n,m} \hat R_{nm}(z)
         e^{2i \pi nx} e^{-2i \pi mx'}
        \ee
and hence for all functions
        $ \rho \in {\cal A}[0,1] $
we have
        \be \label {39}
         \left( R(z) \rho\right)(x) = \intl_{0}^{1}\,R(x,x',z)
         \rho(x')\,dx'=
         \frac1{1-z} \intl_{0}^{1}\rho(x')\,dx' \One(x)
         +{\Sigma_{1}}+{ \Sigma_{2}}+{ \Sigma_{3}},
        \ee
where
        \be \label{40}
         { \Sigma_1} = -\sum_{k\ge 0} \frac1 {2z^{k+1}}
         \sum_{n \ne 0} e^{2i \pi nx}
\left[ \intl_{0}^{1}e^{-2i \pi n 2^k x'} \rho(x')\, dx'
          \right.
          + \left.
        \intl_{0}^{1}e^{2i \pi n2^k x'} \rho(x')\, dx'
        \right] ,
        \ee
        \ba \label{41}
         { \Sigma_2}= - \frac1{2z} \sum_{n \ne 0}
         e^{2i \pi nx}
         \left[
          \intl_{0}^{1}e^{-2i \pi n x'} \rho(x')\, dx' -
          \intl_{0}^{1} e^{2i \pi nx'} \rho(x')\, dx'
          \right ] \, ,
         \ea
         \be \label{42}
         {\Sigma_{3}} = - \sum_{k\ge 0}
         \frac{i}{z^{2}\,z^{k}\pi}
         \sum_{l, n} \frac{e^{2i \pi nx}} {2l-2n+1}
         \left[
         \intl_{0}^{1}e^{-2i \pi (2l+1)2^{k}\,x'}\rho(x')dx'
         \right.         %                      \nonumber \\
         + \left.
         \intl_{0}^{1}e^{2i \pi (2l+1)n2^{k}\,x'}\rho(x') \, dx'
         \right] \, .
        \ee

   Let us calculate the series
        ${ \Sigma_1 , \Sigma_2, \Sigma_3} $
from~(\ref{39}) separately.

   At first, let us consider the integral
        $ \intl_0^1 e^{2i \pi mx} \rho(x)\, dx ,
         \quad m~\in~\Integer . $
Then, integrating
        $ N $
times by parts, we have
        \ba \label {43}
         \intl_{0}^{1}\,e^{2i \pi mx}\rho(x)dx =
         \sum_{s=1}^{N}\frac{(-1)^{s-1}} {(2i \pi  m)^{s}}
         \left[\rho^{(s-1)}(1)\,-\,\rho^{(s-1)}(0)\right]
         %\nonumber \\
         + \frac{(-1)^{N}} {(2i \pi  m)^{N-1}}
         \intl_{0}^{1}\,\rho^{(N)}(x)\,e^{2i \pi mx}\,dx.
        \ea
We note now that in
        $ \Sigma_1 $
and
        $ \Sigma_3 $
one should integrate by parts the expression
        \be \label {44}
         \intl_0^1 \pa{e^{2i \pi mx} + e^{- 2i \pi mx}}
         \rho(x) \, dx
        \ee
where for
        $ \Sigma_1 $
        $ m = n 2^k $
is integer and for
        $ \Sigma_3  $
        $ m = (2l + 1)n 2^k $
is integer. Then, similar to~(\ref{43}) we have
        \ba \label {45}
         \intl_{0}^{1} \left( e^{2i \pi mx} \right.\!\!\!\!&+&\!\!\!\!\left.
         e^{-2i \pi mx}\right) \rho(x)\,dx
         \sum_{l=1}^{N}\frac{(-1)^{l-1}} {(2i \pi  m)^{l}}
         \left[1+(-1)^{l} \right]\cdot
         \left[\rho^{(l-1)}(1)-\rho^{(l-1)}(0)\right] \nonumber \\
        &\!\!\!\! + &\!\!\!\!\frac{(-1)^{N}} {(2i \pi  m)^{N-1}}
         \left[1+(-1)^{N-1} \right]
         \intl_{0}^{1}\,\rho^{(N)}(x)
         \left(e^{2i \pi mx}+e^{-2i \pi mx} \right)\,dx.
        \ea
In the right hand side of~(\ref{45}) the zero terms of the first
sum correspond to odd
        $ l $.
Then, taking
        $ l = 2s $
and replacing
        $ N \to 2N + 1 $
from~(\ref{45}) we obtain
        \ba \label {46}
         \intl_{0}^{1}\,\left( e^{2i \pi mx} \right. \!\!\!\!&+&\!\!\!\! \left.
         e^{-2i \pi mx}\right) \rho(x)\,dx
         = -2 \sum_{s=1}^N \frac1{(2i \pi  m)^{2s}}
       \left[\rho^{(2s-1)}(1)-\rho^{(2s-1)}(0)\right]
         \nonumber \\ \!\!\!\!&-&\!\!\!\!
         \frac2 {(2i \pi  m)^{2N}}
         \intl_0^1 \rho^{(2N+1)}(x)
       \left(e^{2i \pi mx}+e^{-2i \pi mx} \right)\,dx.
        \ea
Let us note that in the expression
        $ \Sigma_2 $
there are the integrals of the type~(\ref{44}), but with
the difference of its components. Consider such integral
and integrating by parts we get
        \ba \label {47}
         \intl_{0}^{1}
         \left( e^{2i \pi mx} \right. \!\!\!\!&-&\!\!\!\! \left. e^{-2i \pi mx}
         \right) \rho(x)\,dx =\sum_{l=1}^{N}
       \frac{(-1)^{l-1}}{(2i \pi  m)^{l}}
       \left[1-(-1)^{l} \right]\cdot
       \left[\rho^{(l-1)}(1)-\rho^{(l-1)}(0)\right] \nonumber
       \\  \!\!\!\!&+& \!\!\!\! \frac{(-1)^{N}} {(2i \pi  m)^{N-1}}
       \left[1-(-1)^{N-1} \right]
       \intl_{0}^{1}\rho^{(N)}(x)
       \left(e^{2i \pi mx}+e^{-2i \pi mx} \right)\,dx.
        \ea
In the right hand side of~(\ref{47}) zero terms of the first sum
correspond to even
        $ l $.
Then, taking
        $ l = 2s + 1 $
and replacing
        $ N \to 2N + 2 $,
from~(\ref{47}) we obtain
        \ba \label {48}
         \intl_{0}^{1}
         \left( e^{2i \pi mx} \right.\!\!\!\!&-& \!\!\!\! \left.
         e^{-2i \pi mx}\right) \rho(x)\,dx =
         % \nonumber \\         &&
         2 \sum_{s=0}^N \frac1{(2i \pi n)^{2s+1}}
         \left[\rho^{(2s)}(1)-\rho^{(2s)}(0)\right]
         \nonumber \\ \!\!\!\!&+&\!\!\!\!
         \frac2 {(2i \pi n)^{2N+1}}
         \intl_0^1 \rho^{(2N+2)}(x)
         \left(e^{2i \pi nx}+e^{-2i \pi nx} \right)\,dx.
        \ea
Then, according to~(\ref{45}),~(\ref{46}) we obtain at
        $ m = n2^k $
the expression for $ \Sigma_1 $:
        \ba \label {49}
         {\Sigma_{1}}&=&
         \sum_{k=0}^{\infty} \frac1{z^{k+1}}
         \sum_{n\ne 0} \sum_{l=1}^{N}
         \frac{e^{2i \pi nx}} {(2i \pi n)^{2l}\,2^{2kl}}
         \left[\rho^{(2l-1)}(1)\,-\,\rho^{(2l-1)}(0)\right]
         \nonumber \\
         &&+ \sum_{k=0}^{\infty}\frac{1}{z^{k+1}}
         \sum_{n\ne 0}
         \frac{e^{2i \pi nx}} {(2i \pi n)^{2N}\,2^{2Nk}}
         \times
         \intl_{0}^{1}\rho^{(2N+1)}(x')
         \left( e^{2i \pi n2^{k}x'}
         + e^{-2i \pi n2^{k}x'}\right)dx'.
        \ea
Now we restrict our consideration by
the set of all densities
        $ \rho(x) $
such that the series in the right hand side of~(\ref{49})
converge when
        $ N \to \infty . $
In the domain
        $ z > 1/2 $
this condition is the condition of the
convergence of the series:
        \be \label {50}
         \sum_l (2i \pi)^{-2l}
         \br{\rho^{(2l-1)}(1) - \rho^{2l-1}(0)} < \infty .
        \ee
The condition
        \be \label {51}
         (2 \pi)^{-2N} \int_0^1 |\rho^{(2n-1)}(x)|\, dx \to 0
         \quad {\rm  at } \quad N \to 0
        \ee
guaranties that the latter term in the right hand side of~(\ref{49})
including the integral converges to zero at
        $ N \to 0. $
So, keeping the condition~(\ref{50}),~(\ref{51}) in
the limit when
        $ N \to \infty $
we obtain for the
        $ \Sigma_1 $
the expression
        \be \label {Sigma1 new}
         \Sigma_1 = \frac1z \sum_{k=0}^\infty \frac1{z^k}
         \sum_{n \ne 0} e^{2i \pi nx} \cdot
         \sum_{l=1}^\infty \frac1{(2i \pi n)^{2l}2^{2k}}
         \br{\rho^{(2l-1)}(1) - \rho^{2l-1}(0) } \, .
        \ee
If
        $ \left| \frac1{z2^{2l}} \right| < 1 $
i.e.
        $ | z 2^{2l} | > 1 $
for any
        $ l, \,\, l \ge 1 $ ,
then the following is valid:
        \begin{eqnarray*} \label {53}
         \frac{1}{z}\,\sum_{k=0}^{\infty}\,
         \frac{1} {(z\,2^{2l})^{k}}=
         \frac{1}{z}\,\left(\frac{1} {1-\frac{1} {z\,2^{2l}}}
       \right)= \frac{1}{z}\,\cdot \frac{z}
         {z-\frac{1}{2^{2l}}}= \frac{1}{z-\frac{1}{2^{2l}}}.
        \end{eqnarray*}
So, if
        $ |z| > 1/4 $
we obtain from~(\ref{Sigma1 new}) and~(\ref{53}):
        \ba \label {54}
         \Sigma_{1} =
         \sum_{l = 1}^{\infty}
         \frac{\left[\rho^{(2l-1)}(1)- \rho^{(2l-1)}(0)\right]}
       {z-(1/2)^{2l}}\,
         \sum_{n\ne 0} \frac{e^{2i \pi nx}}{(2i \pi n)^{2l}} \,.
         \ea
We reconstruct the expression~(\ref{54}) in the following
way:
        \ba \label {55}
         \sum_{n\ne 0}\, \frac{e^{2i \pi nx}}{(2i \pi n)^{2l}}=
         \sum_{n=1}^{\infty}\, \frac{2\cos{(2\pi nx)}}
       {(2\pi n)^{2l}\,(-1)^{l}}.
         \ea
Let us use now the formula for the Bernoulli polynomials~\cite{5}:
        \ba \label {56}
         B_{2l}(x)= \frac{(-1)^{l-1}\,2(2l)!} {(2\pi)^{2l}}\,
         \sum_{n=1}^{\infty}\,\frac{\cos{(2\pi nx)}} { n^{2l}}.
        \end{eqnarray}
Then, the equation~(\ref{55}) can be written as
        \be \label {57}
         \sum_{n \ne 0} \frac{e^{2i \pi nx}}{{2i \pi n}^{2l}} =
         - \frac{B_{2l}}{(2l)!} \, .
        \ee
Taking this into the account we obtain for the
        $ \Sigma_1 $
the following re\-pre\-sen\-ta\-tion
        \be \label {58}
         \Sigma_{1} =
         \sum_{l = 1}^{\infty}
         \frac{\left[\rho^{(2l-1)}(1)- \rho^{(2l-1)}(0)\right]}
         {(2l)!}
         \sum_{n\ne 0} \frac{B_{2l}}{z - (1/2)^{2l}}.
        \ee
Let us consider now the contribution of
        $ \Sigma_2 $
(see~(\ref{41})) into the resolvent~(\ref{39}). For the
calculation $ \Sigma_2 $ we use the formula~(\ref{48}) and obtain
         \ba \label {59}
         {\Sigma_{2}} \!\!\!\!&=&\!\!\!\!\frac{1}{z}\,
         \sum_{n\ne 0}e^{2i \pi nx}
         \,\sum_{l=0}^{N} \frac{1}{(2i \pi n)^{2l+1}}\,
       \left[\rho^{(2l)}(1)\,-\,\rho^{(2l)}(0)\right]\,\\
         \nonumber
         && \!\!\!\! + \frac{1}{z}\,\sum_{n\ne 0}
       \frac{e^{2i \pi nx}}
       {(2i \pi n)^{2N+1}}\,
       \cdot \intl_{0}^{1}\,\rho^{(2N+2)}(x')\,
       \left( e^{2i \pi nx'} + e^{-2i \pi nx'}\right) \,dx'.
        \ea
At the same conditions~(\ref{50}),~(\ref{51}) one can consider a
limit at
        $ N \to \infty $
and as a result we have
        \be \label {60}
         {\Sigma_{2}} = \frac{1}{z}\,
         \sum_{l = 0}^\infty
         \br{\rho^{(2l)}(1)-\rho^{(2l)}(0)}
         \sum_{n \ne 0} \frac{e^{2i \pi nx}}
         {(2i \pi n)^{2l+1}} \, .
        \ee
We rewrite the inner sum in~(\ref{60}) as follows:
        \ba  \label{61}
         \sum_{n\ne 0}\frac{e^{2i \pi nx}} {(2i \pi n)^{2l+1}}=
         \sum_{n=1}^{\infty} \frac{e^{2i \pi nx}-e^{-2i \pi nx}}
       {(2\pi n)^{2l+1}\,(i)^{2l+1}}=
         \sum_{n=1}^{\infty} \frac{(-1)^{l}\,2\sin{(2\pi nx)}}
         {(2\pi n)^{2l+1}}.
         \ea
Once again we use the formula for the Bernoulli polynomials~[5]:
        \ba \label{62}
         B_{2l+1}(x)=\frac{(-1)^{l+1}\,2(2l+1)!} {(2\pi)^{2l+1}}
         \,\sum_{n=1}^{\infty}\, \frac{\sin{(2\pi nx)}}
       { n^{2l+1}},
        \ea
and find that~(\ref{61})~can be written in the form
        \ba  \label{63}
         \sum_{n\ne 0}\frac{e^{2i \pi nx}} {(2i \pi n)^{2l+1}}=
         - \frac{B_{2l+1}}{(2l + 1)!} \, .
        \ea
Then, for the contribution of $ \Sigma_2 $ we have
        \be \label {64}
         \Sigma_2 = -\frac1{z}
         \sum_{l = 1}^{\infty}
         \frac{\br{\rho^{(2l)}(1)- \rho^{(2l)}(0)}}
         {(2l+1)!} B_{2l+1}(x) \, .
        \ee

   Now let us calculate $ \Sigma_3 $.
For this purpose we substitute~(\ref{46}) into~(\ref{42}) at
        $ m = (2l + 1) 2^k $:
        \ba \label {65}
       {\Sigma_{3}} &=&\!\!\!\! \sum_{k \ge 0} \frac{i}{z^2 z^k \pi}
         \sum_{l,n; n \ne 0} \frac{e^{2i \pi nx}}{2l - 2n + 1}
         \sum_{s=1}^{N}
         \frac {\rho^{(2s-1)}(1)-\rho^{(2s-1)}(0)}
         {\br{2i \pi (2l+1)}^{2s}2^{2ks}}  \nonumber \\
         &-& \!\!\!\!\sum_{k=0}^{\infty}\frac{2i}{\pi (2i \pi 2^k)^{2N}}
       \sum_{l,n; n \ne 0} \frac{e^{2i \pi nx}}
         {(2l + 1)^{2N}(2l - 2n + 1)} \times \nonumber \\
         &\times& \!\!\!\! \intl_{0}^{1} \rho^{(2N+1)}(x)
       \left( e^{2i \pi (2l+1)2^{k}x}
         + e^{-2i \pi (2l+1)2^{k}x}\right) \,dx.
        \ea
Since the series over $n$ in the equation~(\ref{65}) does not
converges absolutely one can not take the limit at $ N \to \infty
$ in~(\ref{65}). Therefore in the first term in~(\ref{65}) we
change the places of the last finite sum which depends on $ n $
and the series which depend on $l,n$. Then, the first term
in~(\ref{65}) has the form
        \ba \label{66}
         \sum_{k\ge 0}\,\frac{2}{iz^{2}\,z^{k}}\,
         \sum_{s=1}^{N}\,
       \frac{\rho^{(2s-1)}(1)\,-\,\rho^{(2s-1)}(0)}
       {\left[ (2i\pi)^{2s}\,2^{2ks}\right]}\,
         \sum_{l,n \atop n\ne 0}\, \frac{e^{2i \pi nx}}
       {\left[2n-(2l+1)\right](2l+1)^{s}\,\pi} .
        \ea
We consider the series which depend on $ n $ in~(\ref{66}) and
transform them in the following way:
        \ba  \label {67}
         \sum_{n\ne 0}\, \frac{e^{2i \pi nx}}
       {\left[2n-(2l+1)\right]\,\pi}\, &=&
         \sum_{n}\, \frac{e^{2i \pi nx}}
       {[2n-M]\,\pi}+\frac{1}{(2l+1)\pi}\, =
        % \nonumber \\         &=&
        \sum_{n}\, \frac{e^{2i \pi nx}}
       {2n\pi-M\pi}+\frac{1}{M\pi},
        \ea
where
        $ M = 2l + 1 $ .
We use now the Poisson formula for the summation of the
series~\cite{33}. Let $g(x)$ be a function of the bounded
variation on the interval
        $ (-\infty, +\infty) $
such that the integral in the Poisson summation formula converges
(for example in the sense of the principal value). Then, the
Poisson summation formula has a form
        \be \label {68}
         \sum_{n=-\infty}^{\infty}\,g(2\pi n)=
         \sum_{n=-\infty}^{\infty}\, \frac{1} {2\pi}\,
         \intl_{-\infty}^{+\infty}\,g(t)\, e^{-itn}\,dt.
        \ee
Now we use equation~(\ref{68}) for summation of the
series~(\ref{67}). In our case
        \be \label {69}
         g(t) = \frac{e^{itx}}{t - M \pi},
        \ee
and hence from~(\ref{67}) we get
        \be \label {70}
         \sum_n \frac{e^{2i \pi nx}}{2 \pi n - \pi M} =
         \sum_n \frac1{2 \pi} {\rm V.p.}
         \int_{-\infty}^{+\infty}
         \frac{e^{it(x - n)}}{t - \pi M} \, dt \, .
        \ee
Let us denote
        $ \alpha = x - n $ .
Then, depending on the sign of
        $ \alpha $
one should calculate the integral in~(\ref{70}) by residues
closing the contour into the upper
        ($ \alpha > 0 $)
or into the lower
        ($ \alpha < 0 $)
halfplane. Consider the contour
        $ \Gamma_R $
consisting of two intervals of the real axis:
        $ [-R, -\varepsilon] $
and
        $ [\varepsilon, R] $,
        $ \varepsilon > 0 $;
the halfcircumference
        $ C_\varepsilon $
with the radius
        $ \varepsilon > 0 $
lying in the upper complex halfplane and of the halfcircumference
        $ C_R $
with the radius $R > \varepsilon > 0$ lying also there. Then (the
direction of the clockwise or contra clockwise turning is noted by
arrows):
        \be \label{71}
         \intl_{\Gamma_{R}}=
         \intl_{-R}^{-\varepsilon} + \intl_{\varepsilon}^{R} +
         \intl_{\to C_{\varepsilon}} + \intl_{\gets C_{R}}.
        \ee
One could show that
        \be \label {72}
         \intl_{\gets C_{R}} \to 0 \qquad {\rm if} \quad
         R \to \infty ,
        \ee
and
        \be \label{73}
         \lim_{\varepsilon \to 0}
         \left( \intl_{\to C_{\varepsilon}}\right)\,
         =-i\pi {\rm res} \left\{\frac{e^{it\alpha}}
         {t-\pi M}\right\}_{t=\pi M}.
        \ee
By means of~(\ref{71})~--~(\ref{73}) we obtain for $ \alpha > 0 $:
        \be \label {74}
         {\rm V.p.}\,\intl_{-\infty}^{+\infty}
         \frac{ e^{it \alpha} }
         {t-\pi M} \, dt=
         i\pi {\rm res} \left\{
         \frac {e^{it \alpha}} {t-\pi M} \right\}_{t=\pi M}\,
         = -i\pi e^{i\pi M\alpha}.
        \ee
Quite similar for
        $ \alpha < 0 $, closing the contour into the low
half\-plane, we obtain
        \be \label {75}
         {\rm V.p.} \,\intl_{-\infty}^{+\infty}
         \frac{e^{ith}}{t-\pi M}\,dt=
         -i\pi \,{\rm res}  \left\{
        \frac{e^{ith}}{t-\pi M} \right\}_{t=\pi M}\,
         =\,-i\pi e^{i\pi M\alpha}.
        \ee

   Turning  back to the series~(\ref{67}) one can see that these series are
splited into two series in depending on the sign of
        $ x - n, \quad x \in (0,1) $.
Namely
        \ba \label {76}
         &&x - n > 0 \Longleftrightarrow x > n
         \Longleftrightarrow n = 0, \, -1, \, -2, \, \ldots \\
        % \ee
        % \be \label {77}
         &&x - n < 0 \Longleftrightarrow x < n
         \Longleftrightarrow n = 1, \, 2, \, 3, \,\ldots
        \ea
Hence
        \ba \label{78}
         \sum_{n} \,\frac{1}{2\pi}\,
         \intl_{-\infty}^{+\infty}\,
         \frac{e^{it(x-n)}}
       {t-\pi M}\,dt \!\!\!\!&=&\!\!\!\!
         \frac{i}{2}\,
         \sum_{n=-\infty}^{0}\,e^{i\pi M(x-n)} \,-\,
         \frac{i}{2}\,\sum_{n=1}^{\infty}\,e^{i\pi M(x-n)} \,=
         \nonumber \\
         \!\!\!\!&=&\!\!\!\!\frac{i}{2}\,e^{i\pi M x}  +
         \frac{i}{2}\,\sum_{n=1}^{\infty}\,e^{i\pi M x}\,e^{i\pi Mn}
         % \nonumber \\
         - \frac{i}{2}\,\sum_{n=1}^{\infty}\,e^{i\pi M x}\,e^{-i\pi Mn}.
        \ea
Since for every
        $ M, \, n $
we have
        \be \label {79}
         e^{-i\pi Mn}=e^{i\pi Mn}=
         \left\{
         \begin{array}{rcl}
         +1,\quad Mn\,&=&\,2s\\
         -1,\quad Mn\,&=&\,2s+1.
         \end{array}
         \right.
        \ee
It means that two latter series in~(\ref{78}) are cancelled, and
we obtain
        \be \label {80}
         \sum_{n}\,
         \frac{e^{2i \pi nx}}
       {2\pi n\,-\,\pi M}=
         \sum_{n}\,
         \frac{1}
       {2\pi }\,
         \intl_{-\infty}^{+\infty}\,
         \frac{e^{it(x-n)}}
       {t-\pi M}\,dt=
         \frac{i}{2}\,e^{i\pi M x}, \,\quad\, M=2l+1.
        \ee
However in~(\ref{80}) we are interested in the series of the left
hand side not for all
        $ n $,
but only for
        $ n \ne 0 $.
So, from~(\ref{67}) it follows that
        \ba \label {81}
         \sum_{n\ne 0}\,
         \frac{e^{2i \pi nx}}
       {2\pi n-\pi M}\!\!\!\! &=& \!\!\!\!
         \sum_{n}\,
         \frac{e^{2i \pi nx}}
       {2\pi n-\pi M} +
         \frac{1}
       {\pi (2l+1)}= %\nonumber \\ &=&
         \frac{i}{2}\,e^{i\pi (2l+1) x} + \frac{1}{\pi (2l+1)}.
        \ea
We substitute now this expression in the series depended of $ l $
in~(\ref{66}) and obtain
        \ba \label{82}
         \sum_{l,n \atop n\ne 0}\,
         \frac{e^{2i \pi nx}}
         {\left(2\pi n\,-\,(2l+1)\pi \right)(2l+1)^{2s}}=
         \sum_{l}\,\frac{i}{2}\,
         \frac{e^{i\pi (2l+1)x}}{(2l+1)^{2s}}
         + \sum_{l}\,\frac{1}{\pi\,(2l+1)^{2s+1}}.
        \ea
The latter series in the right hand side in~(\ref{82}) obviously
is equals to zero:
        \be \label {83}
         \sum_l \frac1{\pi (2l+1)^{2s + 1}} = 0 \, .
        \ee
Taking into account relations written above let us consider now
the limit at $ N \to \infty $ in $ \Sigma_3 $. At the same
conditions as above the terms in $ \Sigma_3 $ including the
integrals of $ \rho^{2N + 1} (x) $ tend to zero and so in the
limit $ N \to 0 $ we get
        \be \label{84}
         {\Sigma_{3}}=
         \sum_{k= 0}
         \frac{1}
         {z^{2}\,z^{k}}\,
         \sum_{s=1}^{\infty}
         \frac{\left[ \rho^{(2s-1)}(1)-\rho^{(2s-1)}(0)\right]}
         {(2i \pi )^{2s}\,2^{2ks}}\,
         \sum_{l}\,
         \frac{e^{i \pi (2l+1)x}}
         {(2l+1)^{2s}}.
        \ee
We use the fact that if $ | z | > 1/4 $~(see~(\ref{53})) then
        \be \label {85}
         \sum_{k = 0}^\infty \frac1{(z 2^{2s})^k} =
         \frac{z}{z - (1/2)^{2s}}
        \ee
and for the contribution of
        $ \Sigma_3 $
in the resolvent we obtain the following representation:
        \ba \label {86}
         {\Sigma_{3}}\!\!\!\! &=& \!\!\!\!
         \sum_{s=1}^{\infty}
         \frac{\left[ \rho^{(2s-1)}(1)-\rho^{(2s-1)}(0)\right]}
       {z\left(z-\frac{1}{2^{2s}}\right)(2i \pi )^{2s}}\,
         \sum_{l}\,
         \frac{e^{i \pi (2l+1)x}}
       {(2l+1)^{2s}}\,= \nonumber \\
         &=& \!\!\!\! \sum_{s=1}^{\infty}
         \frac{\left[ \rho^{(2s-1)}(1)-\rho^{(2s-1)}(0)\right]}
       {z\left(z-\frac{1}{2^{2s}}\right)(-1)^{s}(2\pi )^{2s}}\,
         \sum_{l=0}^{\infty}\,
         \frac{e^{i\pi (2l+1)x}+e^{-i\pi (2l+1)x}}
       {(2l+1)^{2s}}\,= \nonumber \\
         &=& \!\!\!\! \sum_{s=1}^{\infty}
         \frac{2\left[ \rho^{(2s-1)}(1)-\rho^{(2s-1)}(0)\right]}
       {z\left(z-\frac{1}{2^{2s}}\right)\,2^{2s}\pi^{2s}}\,
         \sum_{l=0}^{\infty}\,
         \frac{\cos{(2l+1)\pi x}}
       {(2l+1)^{2s}}.
        \ea

Now we use the well-known~\cite{5} Euler polynomials
representation:
        \be \label {87}
         E_{2s-1}(x) = \frac{4(-1)^s(2s-1)!}{\pi^{2s}}
         \sum_{l=0}^\infty
         \frac{ \cos(2l+1) \pi x}{(2l+1)^{2s}}
        \ee
and obtain for
        $ \Sigma_3 $ :
        \be \label {88}
         {\Sigma_{3}}=
         \sum_{s=1}^{\infty}
         \frac{\left[ \rho^{(2s-1)}(1)-\rho^{(2s-1)}(0)\right]}
         {z\,\left(z-\frac{1}{2^{2s}}\right)2^{2s}}\,
         \frac{E_{2s-1}(x)}
         {2(2s-1)!}.
        \ee
To get the desired expression for
        $ \Sigma_3 $
we shall transform the de\-no\-mi\-na\-tor in~(\ref{88}). Namely,
since
        \ba
         \frac{1}
         {z\,\left(z-\frac{1}{2^{2s}}\right)}=
         \pa {\frac{1}
         {z\,-\,\frac{1}{2^{2s}}}\,-
         \,\frac{1}{z}} 2^{2s} \, ,
        \ea
and hence
        \ba \label{90}
         \frac{1}
         {z\,\left(1-\frac{1}{2^{2s}}\right) 2^{2s}}=
         \frac{1}
         {z\,-\,\frac{1}{2^{2s}}}\,-
         \,\frac{1}{z} \,
        \ea
we finally obtain for $ \Sigma_3 $
        \ba \label {91}
         {\Sigma_{3}} =\sum_{s=1}^{\infty}
         \frac{\left[ \rho^{(2s-1)}(1)-\rho^{(2s-1)}(0)\right]}
         {\left(z-\frac{1}{2^{2s}}\right)(2s-1)!}\,
         \cdot
         \frac{E_{2s-1}(x)}{2}
         %\nonumber \\
         - \frac{1}{z}\,
         \sum_{s=1}^{\infty}
         \frac{\left[ \rho^{(2s-1)}(1)\,-\,\rho^{(2s-1)}(0)\right]}
         {(2s-1)!}\,
         \cdot\,\frac{E_{2s-1}(x)}{2} \, .
        \ea

   Now we return again to the presentation~(\ref{39}) of the
resolvent. As a result of the above calculations we find the
expressions for the contributions of
        $ \Sigma_1 $,
        $ \Sigma_2 $,
        $ \Sigma_3 $
in the terms of the Bernoulli and Euler polynomials. We sum this
contributions according to~(\ref{39}) and collect the terms at the
different peculiarities by
        $ z $.
Namely collecting terms at~$1/z$ according~(\ref{91}),~(\ref{64})
we obtain
        \ba \label {92}
         -\frac{1}{z}\,\sum_{l=0}^{\infty}
         \frac{\left[\rho^{(2l)}(1)-\rho^{(2l)}(0)\right]}
       {(2l+1)!}
         B_{2l+1}(x)
         %\nonumber \\
         \!\!\!\!&-&\!\!\!\! \frac{1}{z}\,\sum_{l=0}^{\infty}
         \frac{\left[\rho^{(2l+1)}(1)-\rho^{(2l+1)}(0)\right]}
       {(2l+1)!(2l+2)}
         \frac{E_{2l-1}(x)(2l+2)}{2}= \nonumber \\
         =-\frac{1}{z}
         \sum_{l=0}^{\infty}
         \frac{\left[\rho^{(2l)}(0)-\rho^{(2l)}(1)\right]}
       {(2l+1)!}
         B_{2l+1}(x)
         % \nonumber \\
         \!\!\!\!&+&\!\!\!\! \frac{1}{z}\,\sum_{l=0}^{\infty}
         \frac{\left[ \rho^{(2l+1)}(0)-\rho^{(2l+1)}(1)\right]}
       {(2l+2)!}
         (l+1) E_{2l+1}(x)= \nonumber\\
         &=& \!\!\!\!-\frac{1}{z}\,
         \sum_{m=0}^{\infty}
         \frac{\left[ \rho^{(m)}(1)-\rho^{(m)}(0)\right]}
       {(m+1)!}\,
         \psi_{m}(x).
        \ea
Here
        \be \label {93}
         \psi_{m}(x)=
         \left\{
         \begin{array}{lcl}
         B_{2l+1}(x),             && m = 2l\\
         (l+1)E_{2l+1}(x),     && m = 2l+1.
         \end{array}
         \right.
        \ee
We collect now the terms at the pole peculiarities
        $ \pa{z - \frac1{2^{2s}}}^{-1} $
in the resolvent~(\ref{39}) using~(\ref{88}), (\ref{58}):
        \ba \label {94}
         \sum_{l=1}^{\infty}
         \frac{\left[ \rho^{(2l-1)}(0)-\rho^{(2l-1)}(1)\right]}
       {z\,-\,\frac{1}{2^{2l}}} \!\!\!\! & \cdot & \!\!\!\!
         \left\{
       \frac{B_{2l}(x)}
       {(2l)!}-
       \frac{E_{2l-1}(x)}
       {2(2l-1)!}\right\}= \nonumber \\
         &=& \!\!\!\!\sum_{l=1}^{\infty}
         \frac{\left[ \rho^{(2l-1)}(1)-\rho^{(2l-1)}(0)\right]}
       {z-\frac{1}{2^{2l}}}
         \varphi_{2l-1}(x),
        \ea
where
        \be \label {95}
         \varphi_{2l-1}(x) = -l E_{2l-1}(x) + B_{2l}(x).
        \ee
From~(\ref{92})~--~(\ref{95}) according to~(\ref{39}) we obtain
the action of the resolvent $ R(z) $ in the form
        \ba \label{96}
         \left(R(z)\rho\right)(x)\!\!\!\!&=& \!\!\!\!
         \frac{1}{1-z}\,\intl_{0}^{1}\,\rho(x')\,dx'\, \One(x)
         %\nonumber \\
         + \sum_{l=1}^{\infty}
         \frac{\left[ \rho^{(2l-1)}(0)-\rho^{(2l-1)}(1)\right]}
       {\left(z-\frac{1}
       {2^{2l}}\right)(2l)!}
         \varphi_{2l-1}(x)\, \nonumber \\
         &&\!\!\!\!+\frac{1}{z}\,\sum_{m=0}^{\infty}
         \frac{\left[\rho^{(m)}(0)\,-\,\rho^{(m)}(1)\right]}
       {(m+1)!}\,
         \psi_{m}(x).
        \ea
This formula (as one for the Renyi map~\cite{1},~\cite{4})
determine the continuation
        $ R^c (z) $
of the resolvent
        $ R(z) $
of the operator
        $ U_T $
to the whole complex plane
        $ \Complex $
of the spectral parameter
        $ z $.
The continuation
        $ R^c (z) $
needs the restriction
        $ U_T $
to the suitable functional space
        $ \Phi \subset {\cal A}[0,1] $. In the examining case
the choice of
        $ \Phi $
is caused by the reasons of the rightness of the developing above
calculations, i.~e. the functions $\rho (x)$, $\rho \in \Phi $
should satisfy the conditions~(\ref{50}),~(\ref{51}). As for the
Renyi map~\cite{1},~\cite{4}, the detailed description of the test
space can be found in~\cite{16}.

  It follows from~(\ref{96}) that the continuation
        $ R^c (z) $
of the restricted re\-sol\-vent
        $ R(z)|_\Phi $ can be written in the form
        \be \label{97}
         R^{c}(z)=\frac{1}{1-z} \, |\One\rangle\langle\One| +
         \frac{1}{z}\,\sum_{m=0}^{\infty}\,| \psi_{m}\rangle
         \langle\tilde\psi_{m}|\,
         +\,\sum_{l=1}^{\infty}\,
         \frac{1}{z-\frac{1} {2^{2l}}}\,
         |\,\varphi_{2l-1}\rangle\langle\tilde \varphi_{2l-1}|,
        \ee
where the functions
        $ \psi_m $
and $ \varphi_{2l-1} $ are determined by~(\ref{93}) and~(\ref{95})
res\-pec\-tively. By $ \langle\tilde \psi_{m} | $ and
        $ \langle\tilde \varphi_{2l-1}| $
are denoted functionals over the
        $ \Phi $
(i.~e. the elements of the dual space
        $ \Phi^\times $) which are acting to
        $ \rho \in \Phi \subset {\cal A}[0,1] $
according to the rules
        \ba \label{98}
         &&\langle \tilde \psi_{m}| \rho \rangle=
         \frac{\left[\rho^{(m)}(0)-\rho^{(m)}(1)\right]}
         {(m+1)!} \, ,
        \ea
        \ba \label{99}
         &&\langle \tilde \varphi_{2l-1}| \rho \rangle=
         -\,\frac{\left[\rho^{(2l-1)}(1)-\rho^{(2l-1)}(0)\right]}
       {(2l)!}.
        \ea
It follows from these formulas that the functionals
        $ \langle\tilde \psi_{m} | $
and
        $ \langle\tilde \varphi_{2l-1}| $
can be represented in the form of the derivatives of
$ \delta $-functions:
        \ba \label{P100}
         \psi_{m}=
         (-1)^{m}\,\frac{\left[\delta^{(m)}(x)\,-\,\delta^{(m)}
         (x-1)\right]}{(m+1)!}, \\
            \label{P101}
         \varphi_{2l-1}=
         -\,\frac{\left[\delta^{(2l-1)}(x-1)\,-\,\delta^{(2l-1)}(x)\right]}
       {(2l)!},
        \ea
 Coming back to the representation(\ref{97}), it should be
checked, that
        $ \varphi_{2l-1} (x) $
satisfy the equation
        \be \label {102}
         U_T \varphi_{2l-1}(x)=\frac1{2^{2l}} \varphi_{2l-1} (x) \, ,
        \ee
or, what is the same, the equation
        \be \label {103}
         \frac12 \br{\varphi_{2l-1} \pa{\frac{x}2} +
         \varphi_{2l-1} \pa{1 - \frac{x}2}} =
         \frac1{2^{2l}} \varphi_{2l-1} (x) \, ,
        \ee
where
        $ \varphi_{2l-1} (x) $
is given by(\ref{95}). To prove~(\ref{103}) we use the known
formulas~\cite{5}:
        \be \label{104}
         B_{2l}\left(\frac{x}{2}\right)=
         (-1)^{2l}\,B_{2l}\left(1\,-\,\frac{x}{2}\right)=
         B_{2l}\left(1\,-\,\frac{x}{2}\right),
        \ee
        \be \label{105}
         E_{2l-1}\left(\frac{x}{2}\right)\,
         =\,(-1)^{2l-1}\, E_{2l-1}\left(1-\frac{x}{2}\right)=
         -E_{2l-1}\left(1-\frac{x}{2}\right) , \,
        \ee
        \be \label{106}
         E_{n-1}(x)=
         \frac{2}{n}
         \left\{B_{n}(x)\,-\,2^{n}\,B_{n}\left(\frac{x}{2}
         \right)\right\}.
        \ee
Then, by means of~(\ref{104})~--~(\ref{106}), we have
        \ba      \label{107}
         \varphi_{2l-1}\left(\frac{x}{2}\right)+
         \varphi_{2l-1}\left(1-\frac{x}{2}\right)\!\!\!\!&=&\!\!\!\!
         B_{2l}\left(\frac{x}{2}\right)+
         B_{2l}\left(1-\frac{x}{2}\right)
         % \nonumber \\
         - l E_{2l-1}\left(\frac{x}{2}\right)-
         l E_{2l-1}\left(1-\frac{x}{2}\right) = \nonumber \\
         &=&\!\!\!\! 2 B_{2l}\left(\frac{x}{2} \right).
        \ea
Hence to prove~(\ref{103}), it is necessary to be sure that

        \be \label {108}
         B_{2l}(\frac{x}{2}) = \frac{1}{2^{2k}}[B_{2l}(x) - l E_{2l-1}(x)].
        \ee
It follows from~(\ref{108}) that
        \be \label {109}
         l E_{2l-1}(x) = B_{2l}(x) - 2^{2l} B_{2l}
         \pa{\frac{x}2} \, .
        \ee
Substituting this relation to the right hand side of~(\ref{108})
we obtain the identity
        $$
         B_{2l}\left(\frac{x}{2}\right)=
         \frac{1}{2^{2l}}
         \left\{\,B_{2l}(x)\,-\,B_{2l}(x)\,-\,
       2^{2l}\,B_{2l}\left(\frac{x}{2} \right)\right\}
       =B_{2l}\left(\frac{x}{2} \right)
      $$
which proves~(\ref{103}) and hence~(\ref{102}). It means that
        $ \varphi_{2l-1} (x)$
are the eigenfunctions of the Frobenius--Perron operator
        $ U_T |_\Phi $ ,
restricted to the test functions space
        $ \Phi \subset {\cal A}[0,1] $ .

  Let us mention that $ \varphi_{2l-1} (x) $
can be expressed in the terms of even Bernoulli polynomials.
Namely using~(\ref{109}) we see that
        \be \label {110}
         \varphi_{2l-1} (x) = B_{2l} (x) - l E_{2l-1}(x) =
         2^{2l} B_{2l} \pa{\frac{x}2} .
        \ee
Similarly one can check that the functions
        $ \psi_m (x) \in \Phi \subset {\cal A} [0,1] $ ,
given by the equation~(\ref{93}), are included in
        $ {\rm ker } U_T |_\Phi $
i.~e. satisfy the equation
        \be \label {111}
         (U_T \psi_m)(x) = 0 \, , \quad m \in \Integer .
        \ee
Therefore they are the eigenfunctions of
        $ U_T |_\Phi $
corresponding to the eigenvalue $ z = 0 $ of the infinite
degeneracy. Turning again to the~(\ref{97}) we determine the
extended resolvent
        $ \tilde R_V (z) $
of the Koopman operator
        $ V_T $
for the tent map by duality, which is defined in the Introduction.
Then on the base of~(\ref{97}) we have for
        $ \tilde R_V (z) $ :
        \ba \label {112}
         \tilde R_{V}(z)=\frac{1}{1-z}|\,\One\rangle\langle\One| +
         \frac{1}{z}\,\sum_{m=0}^{\infty}\,| \tilde\psi_{m}\rangle
         \langle\psi_{m}|\,
         %\nonumber \\
         +\,\sum_{l=0}^{\infty}\,
         \frac{1}
         {z-\frac{1}{2^{2l}}}\,
         |\tilde \varphi_{2l-1}\rangle\langle\varphi_{2l-1}|.
        \ea
Now we simplify the notations in~(\ref{98}),~(\ref{99}), taking
into account that the functionals
        $ \langle\tilde \psi_{m} | $
and
        $ \langle\tilde \varphi_{2l-1} | $
act to
        $ \rho \in \Phi $
equally. Namely, we determine the functional
        $ \langle \tilde \chi_{m} |  $
by its action on
        $ \rho \in \Phi $
as follows
        $$
         \langle \tilde \chi_{m} | \rho \rangle =
         \frac{\rho^{(m)}(1)\,-\,\rho^{(m)}(0)}{(m+1)!}.
        $$
Then
         \begin{eqnarray*}
         \langle\tilde \psi_{m} | =\langle\tilde \chi_{m} |, \qquad
         \langle\tilde \varphi_{2l-1} | =\langle\tilde \chi_{2l-1} |.
        \end{eqnarray*}
Using these notations and~(\ref{93}),~(\ref{95}) we can
rewrite~(\ref{97}) in the form:
        \ba \label{113}
         -\,R^{c}(z)= \frac{1}{z-1}|\,\One\rangle\langle\One|
         %\nonumber \\
         \!\!\!\! &+& \!\!\!\! \frac{1}{z}\,\left(
         \sum_{l=0}^{\infty}\,| B_{2l+1}\rangle\langle\tilde\chi_{2l}|\,
         +\,\sum_{l=1}^{\infty}\,|\,l\,E_{2l-1}\rangle\langle\tilde
       \psi_{2l-1}|
         \right)
         \nonumber \\
         &+&\!\!\!\! \sum_{l=1}^{\infty}\,\frac{1}
       {z-\frac{1}{2^{2l}}}\,
         |\,2^{2l}\,B_{2l}
         \left(\frac{x}{2}\right)\rangle
         \langle\tilde\chi_{2l-1}|.
        \ea
We check now the completeness for
        $ R^c (z) $
in the strong sense on the test functions space $\Phi$
using the fact~\cite{1},~\cite{16},~\cite{20} that
        $ \Phi $
coincides with the space of the entire functions of exponential
type less then
        $ 2 \pi $.
For this purpose the following equation should be checked:
        \be \label{114}
         \rho= -\,\sum_{l=1}^{\infty}
         {\rm res} \!\!\!\! \left. \phantom{^2_2} \right|_{z=2^{-2l}} R^{c}
         \rho\,-\,
         {\rm res} \!\!\!\! \left. \phantom{^2_2}\right|_{z=1} R^{c}\,\rho\,-
         \,{\rm res} \!\!\!\! \left. \phantom{^2_2}\right|_{z=0} R^{c}\,
         \rho\,
         \,\quad\,\forall\,\rho\,\in\,\Phi,
        \ee
Proceeding from~(\ref{113}), the right hand side of~(\ref{114}) is
equals to
        \ba \label {115}
         |\,\One\rangle\langle\One|\,\rho\rangle +
         \sum_{l=0}^{\infty}\,| B_{2l+1}\rangle\langle\tilde\chi_{2l}|\,
         \rho\rangle\,
         % \nonumber \\
         \!\!\!\! &+& \!\!\!\! \sum_{l=1}^{\infty}\,|\,l\,E_{2l-1}
         + 2^{2l}\,B_{2l} \left(\frac{x}{2}\right)\rangle
         \langle\tilde\psi_{2l-1}| \rho \rangle\,=\\
         \nonumber
         &=&\!\!\!\! |\,\One\rangle\langle\One|\,\rho\rangle +
         \sum_{n=1}^{\infty}\,|\,B_{n}\rangle\langle\tilde\chi_{n-1}|\,
         \rho\rangle.
        \ea
Here the equation~(\ref{110}) is used as well as the summation
over even and odd indices. Then~(\ref{115}) can be written also in
the form:
        \be      \label{116}
         \int \limits_{0}^{1}\,\rho (x)\,dx +
         \sum_{n=1}^{\infty}\,\frac{\rho^{(n-1)}(1)-\rho^{(n-1)}(0)}
         {n!}\,B_{n}(x)=\rho(x).
        \ee
Here the equality of the left hand side to the function
        $ \rho \in \Phi $
follows from the Euler--Maclaren formula (see~\cite{20}) for
example). Hence, the completeness for
        $ R^c (z) $
and so for the extension of the Koopman operator
        $ \tilde R_V (z) $
is proved.

   It should be noted, that the property of the
completeness can be written formally in the operator form:
        \be      \label {117}
         {\bf I} = |\,\One\rangle\langle\One| +
         \sum_{n=1}^{\infty}\,|\,B_{n}\rangle\langle\tilde\chi_{n-1}|\, .
        \ee
Let us introduce the operators
        \be \label {118}
         \Pi_n = | B_n\rangle \langle \tilde \chi_{n-1} |
        \ee
on
        $ \Phi \subset {\cal A}[0,1] $
and check that
        $ \Pi_n $
satisfies the following conditions
        \be      \label {119}
         \Pi_{l}\,\Pi_{m}=\delta_{l\,m}\,\Pi_{m},\qquad
         \Pi_{0}\,\Pi_{m}=0,
        \ee
where
        $ \Pi_0 $
is determined above by~(\ref{37}). The property~(\ref{119}) means
that $ \{ \Pi_n \} $ is the set of the orthoprojections. We shall
use the fact that the Bernoulli polynomials belong to the family
of the Appel polynomials~\cite{5},~\cite{20} and hence the
following equation is valid:
        \be \label {120}
         B'_n = n B_{n-1} (x)
        \ee
(here the prime means the derivative by $ x $).
From~(\ref{120}) it follows that
        \be \label {121}
         B_m^{(l-1)} (x) = \frac{m!} {(m-l+1)!} B_{m-l+1}(x) \, .
        \ee
It is well known~\cite{5} also that
        \ba \label {122}
         && B_{2n} (0) = B_{2n} (1) \, ,\nonumber \\
         && B_{2n+1} (0) = B_{2n+1} (1) =0 \quad n \ge 1 \, ,\nonumber \\
         && B_1 (0) = -1/2 \, , \nonumber \\
         && B_1 (1) = 1/2 \, .
        \ea

On the base of~(\ref{121}) and~(\ref{122}) we conclude that the
equation
        \be \label {123}
         \frac{B^{(l-1)}_{m}(1)\,-\,B^{(l-1)}_{m}(0)}{l!}=
         \delta_{l m}\,
         \left[\,B_{1}(1)\,-\,B_{1}(0)\right]=
         \delta_{l m}
        \ee
is valid. It should be noted that
        \be      \label {124}
         \langle\tilde\chi_{l-1} | B_{m}\rangle=
         \frac{B^{(l-1)}_{m}(1)\,-\,B^{(l-1)}_{m}(0)}
         {l!}=\delta_{l m}.
        \ee
Then
        $$
         \Pi_{l}\,\Pi_{m}=| B_{l}\rangle\langle\tilde\chi_{l-1}| B_{m}\rangle
         \langle\tilde\chi_{m-1}|=\delta_{l\,m}\,\Pi_{m},
        $$
and so the first of the equations~(\ref{119})  is proved. Now we
should check the second equation in~(\ref{119}) i.~e. we should
calculate
        \be \label {125}
         \Pi_{0}\,\Pi_{m}=| \One \rangle \langle \One | B_{m} \rangle
         \langle\tilde\chi_{m-1}| \, .
        \ee
Again using~(\ref{122}) we find
        \be \label{126}
         \langle\One| B_{m}\rangle=\int \limits_{0}^{1}\,B_{n}(x)\,dx=
         \frac{B_{m+1}(1)\,-\,B_{m+1}(0)}{(m+1)!}=0,
        \ee
and hence the second equation is proved.

   Now let us act by the operator $ U_T  $
to the decomposition~(\ref{117}) from the left. Then taking into
account the properties
        \ba      \label {127}
         && \tilde U_{T}| B_{2l+1}\rangle\,= \,0, \nonumber \\
         \nonumber
         && \tilde U_{T}| B_{2l}\rangle\,=\,
         \frac{1}{2^{2l}}| B_{2l}\,-\,l E_{2l-1}\rangle , \\
          && \tilde U_{T}| \One\rangle\,= \,| \One\rangle
        \ea
which follow from~(\ref{107}),~(\ref{108}),~(\ref{111}), we obtain
the generalized spectral decomposition of the Frobenius--Perron
operator
        $ \tilde U_T  $
corresponding to the extended resolvent
        $ R^c (z) $ :
        \be  \label {128}
         \tilde U_{T}=| \One\rangle\langle\One | +
         \sum_{l=1}^{\infty}\,\frac{1}{2^{2l}}| B_{2l}\,-\,
         l\,E_{2l-1}\rangle\langle\tilde\chi_{2l-1}|.
        \ee
This is the main result of the work. It can be rewritten using the
projections which are different from ones in~(\ref{118}). Namely
let us determine the operators
        \be \label {129}
         \hat \Pi_l = | B_{2l} - l E_{2l+1} \rangle
         \langle \tilde \chi_{2l-1} |
        \ee
and rewrite~(\ref{128}) in the form
        \be \label {130}
         \tilde U_{T}=\Pi_{0} +
         \sum_{l=1}^{\infty}\,\frac{1}{2^{2l}}\hat \Pi_{l}.
        \ee
We shall check that operators
        $ \hat \Pi_l $
are the orthoprojections again, i.~e.
        \be \label {131}
         \hat \Pi_{l}\,\hat \Pi_{m}=
         \delta_{l\,m}\,\hat \Pi_{m},\,\qquad\, \Pi_{0}\,
         \hat \Pi_{l}=0.
        \ee
Indeed from~(\ref{11}) it follows that
        \be \label {132}
         \hat \Pi_{l}=2^{2l}| B_{2 l}\left(\frac{x}{2}\right)\rangle
         \langle\tilde \chi_{2l-1}\,|,
        \ee
and in order to prove~(\ref{131}) it is enough to verify that
        \be  \label {133}
         \langle\tilde \chi_{2l-1}| B_{2 m}\left(\frac{x}{2}\right)\rangle
         = \delta_{l\,m}\,\frac{1}{2^{2l}}.
        \ee
The latter is checked by the direct calculation:
        \ba  \label {134}
         \langle\tilde \chi_{2l-1}| B_{2 m}\left(\frac{x}{2}\right)\rangle
         \!\!\!\! &=& \!\!\!\!
         \frac{1}{2^{2l}}
         \frac{B^{(2l-1)}_{2m}(1/2)-B^{(2l-1)}_{2m}(0)}
         {(2l)!}=
         % \nonumber \\ &=&
         \frac{2m!}{2^{2l}}\cdot
         \frac{B_{2m-2l+1}(1/2)-B_{2m-2l+1}(0)}
         {(2l)!(2m-2l+1)!}\,= \nonumber \\
         &=& \!\!\!\! \frac{1}{2^{2l}}\,\delta_{l\,m}\,
         \left[ B_{1}\left(\frac{1}{2}\right)\,-\,B_{1}(0)\right]=
         \frac{1}{2^{2l}}\, \delta_{l\,m},
        \ea
where we use the properties of the Bernoulli polynomials:
        \ba \label {135}
         && B_{2n+1} (0) = B_{2n+1} (1/2) = 0, \quad n \ge 1
           \nonumber \\
         && B_1 (0) = -1/2  \nonumber \\
         && B_1 (1/2) = 0
        \ea

The second equation in~(\ref{131}) is also can be obtained
directly by using the properties of the Euler polynomials:
        \be \label {136}
         \Pi_{0}\,\hat \Pi_{l}=| \One\rangle\langle\One| B_{2 l}\,
         -\,l\,E_{2l-1}\rangle
         \langle\tilde\chi_{2l-1}\,|.
        \ee
According to~(\ref{126})
        $ \langle \One | B_m \rangle = 0 $  for all $ m $
and so we have to calculate
        \be \label {137}
         \langle\One| E_{2l-1}\rangle=
         \int \limits_{0}^{1}\,E_{2l-1}(x)\,dx=
         \frac{E_{2l}(1)\,-\,E_{2l}(0)}
         {(2l)!}.
        \ee
Using the fact that
        $ E_{2l} (1) = E_{2l} (0) $
we get sure that
        $ \Pi_0 \cdot \hat \Pi_l = 0 $ .

   We have proved above (see~(\ref{102}),~(\ref{103})) that
the original operator $ U_T  $ acts to the right eigenfunctions
"correctly". Now we should check that the extended operator
        $ \tilde U_T  $
given by the generalized spectral decomposition~(\ref{130}) in the
Gelfand triplet
        $ \Phi \subset L_2 (0,1) \subset \Phi^\times $
acts in its spectral representation in the following way:
        $$ \begin{array}{lcl} \label {138}
         \tilde U_{T}| B_{2m}\left(\frac{x}{2}\right)\rangle
         = \frac{1}{2^{2m}}| B_{2m}\left(\frac{x}{2}\right)\rangle,
         &&
         \tilde U_{T}| \One\,\rangle = | \One\rangle,
         \\
         \tilde U_{T}| B_{2l+1}\rangle\,=\,0 , &&
         \tilde U_{T}| E_{2l-1}\rangle\,=0 .
        \end{array}
        $$
Indeed from~(\ref{132}),~(\ref{133}) it follows that
        \be \label {139}
         \hat \Pi_{l}| B_{2m}\left(\frac{x}{2}\right)\,\rangle=
         \delta_{ml}\frac{1}{2^{2m}}| B_{2m}\left(\frac{x}{2}\right),
        \ee
and from~(\ref{136}),~(\ref{137}) it follows that
        \be \label {140}
         \Pi_{0}| B_{2m}\left(\frac{x}{2}\right)\,\rangle\,=0, \,\qquad\,
         \Pi_{0}| E_{2l-1}(x)\,\rangle = 0.
        \ee
Consequently from the equation~(\ref{126}) one can obtain
        \ba \label {141}
         \Pi_{0} \, | B_{2m+1}(x)\rangle\,=0 , \\
            \label {142}
         \hat \Pi_{l}| B_{2m-1}(x)\,\rangle\,=0.
        \ea
Finally from the fact~\cite{20} that the Euler polynomials belong
to the set of the Appel polynomials also (see~\ref{121}) and from
the property
        $ E_{2n} (1) = E_{2n} (0) $
it follows that
        \be \label {143}
         \hat \Pi_l | E_{2m-1} (x)\rangle = 0 \, .
        \ee
Collecting the formulas~(\ref{139})--(\ref{143}) we
obtain~(\ref{138}).

\section{Acknowledgements}
 The authors acknowledge RFBR Grant \# 99-01-00696 for support of this
 work.


\begin{thebibliography}{99}

\bibitem {1} Antoniou I., Dmitrieva L.A., Kuperin Yu.A., Melnikov Yu.B.
Resonances and the Extension of Dynamics to Rigged Hilbert  Spaces
// Computers Math.Applic., vol.34, N 5/6, 1997, 399-425.

\bibitem {2} Driebe D., Ordonez G., Using Symmetries of the Frobenius-Perron
Operator to Determine Spectral Decomposition // Phys.Lett.
A.Vol.211, 1996, 204-210.

\bibitem {3} Fechbah H., Unified Theory of Nuclear Reactions 1, 11 // Ann.Phys.
Vol.5, 1958 357-390; Ibid. vol.19 1962, 287-313.

\bibitem {4} Antoniou I., Tasaki S.,  Spectral Decomposition of the Renyi Map //
J.Phys.A: Math.Gen. (1993), vol.26, 73-94.

\bibitem {5} Abramowitz M., Stigun I., Handbook of Mathematical Functions //
National Bureau of Standards, Applied Mathematics Series - 55, 1964.

\bibitem {6} Schuster H.G., Deterministic Chaos //
Physik-Verlag, Weinheim, 1984.

\bibitem {7} Mackey M.C. Time's Arrow: The Origins of Thermodynamic Behavior //
Springer-Verlag, Berlin-Heidelber-New York, 1993.

\bibitem {8} Prigogin I., Stengers I., Time, Chaos, Quantum // Moscow,"Progress", 1994, á.159.

\bibitem {9} Nikolsckii N. K., Lectures on Shift Operator //
Moscow , "Nauka", 1980.

\bibitem {10} Moren K. Methods of Hilbert Space // Moscow , "Mir", 1965, p.372.

\bibitem {11} Birman M. Ch., Solomyak M. Z., Spectral Theory of Self-adjoint Operators in the Hilbert Spaces // Leningrad, "Izdatelstvo LGU", 1980.

\bibitem {12} Prigogine I. Noneqilibrium Statistical Mechanics, Wiley, New York, 1962.

\bibitem {13} Gelfand I., Vilenkin N., Generalized Functions,
vol.4. Academic Press, New York, 1994.

\bibitem {14} Bohm A., Gadella, Dirac Rets, Gamow Vectors and Gelfand Triplets // Springer Lecture Notes in Physics, vol.348, Springer-Verlag, Berlin, 1989.

\bibitem {15} Lasota A., Mackey M., Probabilistic Properties of Deterministic
Systems, Cambridge Univ. Press, Cambridge, U.K., 1985.

\bibitem {16} Antiniou I., Tasaki S. Generalized Spectral Decomposition of Mixing Dynamical Sys\-tems // Int.J. of Quantum Chemistry, vol.46, 1993, 425-474.

\bibitem {20} Boas K., Buck R., Polynomial Expansions of Analytic Functions,
Springer-Verlag, Berlin, 1958.

\bibitem {21} Hasegawa H., Driebe D., Intrinsic Irreversibility and Validity of
the Kinetic Description of Chaotic Systems // Phys.Rew.E, vol.50, No.3 (1994), 1781--1809.

\bibitem {22} Ruelle D., Phys.Rew.Lett. vol.56, 1986, p.405.

\bibitem {23} Pollicott M., Ann.Math., vol.131, 1990, p.331.

\bibitem {24} Isola S., Commun. Math.Phys. V.116, 1988, p.343.

\bibitem {25} Dorfle M., Spectrum and Eigenfunctions of the Frobenius-Perron
Operator of the Tent Map // J.Stat.Phys. vol.40, N 1/2 1985,
93-132.

\bibitem {26} Danford N., Schwarts J.T., Linear Operators, vol.III, Wiley, New
York, 1971.

\bibitem {27} Prigogine I., From Being to Becoming // Free Man, San Francisco,
1980.

\bibitem {28} Dmitrieva L.A., Guschin D.D., Kuperin Yu.A.,
Generalized Spectral
Analysis of Some Exactly Solvable Chaotic Maps, Proc.Int.Seminar
Day on Diffraction,  St.Petersburg, June 1-3, 1999, p.122-129.

\bibitem {29} Gohbert I. S., Krein M. G., Theory of Volterra
Operators in the Hilbert Space and its Applications // Moscow, "Nauka", 1967.

\bibitem {30} Christiansen F., Paladin G., Rugh H.H., Phys.Rev.Lett., vol.65,
1990, 2087.

\bibitem {31} Artuso R., Phys.Lett.A.,vol.160, 1991, p.528.

\bibitem {32} Gaspard P., J.Stat.Phys., vol.68, 1992, p.68.

\bibitem {33} Encyclopedia of Mathematics (Ed. by  Vinogradov I.
M.), vol.4 // Moscow, "Sovietskaya Entsiklopediya", 1984.

\end{thebibliography}
\end{document}